\documentclass[aps,twocolumn,groupedaddress,showpacs]{revtex4}

\usepackage{graphicx}

\begin{document}

\bibliographystyle{apsrev}

\title{Electronically competing phases and their magnetic field dependence in
electron-doped nonsuperconducting and superconducting
Pr$_{0.88}$LaCe$_{0.12}$CuO$_{4\pm\delta}$}

\author{H. J. Kang}
\affiliation{Department of Physics and Astronomy, The University
of Tennessee, Knoxville, Tennessee 37996-1200, USA}
\author{Pengcheng Dai}
\email{daip@ornl.gov} \affiliation{Department of Physics and
Astronomy, The University of Tennessee, Knoxville, Tennessee
37996-1200, USA} \affiliation{Condensed Matter Sciences Division,
Oak Ridge National Laboratory, Oak Ridge, Tennessee 37831-6393,
USA}
\author{H. A. Mook}
\affiliation{Condensed Matter Sciences Division, Oak Ridge
National Laboratory, Oak Ridge, Tennessee 37831-6393, USA}
\author{D. N. Argyriou}
\affiliation{Hahn-Meitner-Institut, Glienicker Str 100, Berlin
D-14109, Germany}
\author{V. Sikolenko}
\affiliation{Hahn-Meitner-Institut, Glienicker Str 100, Berlin
D-14109, Germany}
\author{J. W. Lynn}
\affiliation{ NIST Center for Neutron Research, National Institute
of Standards and Technology, Gaithersburg, Maryland 20899, USA}
\author{Y. Kurita}
\affiliation{Central Research Institute of Electric Power
Industry, Komae, Tokyo 201-8511, Japan}
\author{Seiki Komiya}
\affiliation{Central Research Institute of Electric Power
Industry, Komae, Tokyo 201-8511, Japan}
\author{Yoichi Ando}
\affiliation{Central Research Institute of Electric Power
Industry, Komae, Tokyo 201-8511, Japan} \altaffiliation[Also at
]{Department of Physics, Tokyo University of Science, Shinjuku-ku,
Tokyo 162-8601, Japan.}

\date{\today}

\begin{abstract}
We present comprehensive neutron scattering studies of
nonsuperconducting and superconducting electron-doped
Pr$_{0.88}$LaCe$_{0.12}$CuO$_{4\pm\delta}$ (PLCCO). At zero field,
the transition from antiferromagnetic (AF) as-grown PLCCO to
superconductivity without static antiferromagnetism can be
achieved by annealing the sample in pure Ar at different
temperatures, which also induces an epitaxial (Pr,La,Ce)$_2$O$_3$
phase as an impurity. When the superconductivity first appears in
PLCCO, a quasi-two-dimensional (2D) spin-density-wave (SDW) order
is also induced, and both coexist with the residual
three-dimensional (3D) AF state. A magnetic field applied along
the $[\bar{1},1,0]$ direction parallel to the CuO$_2$ plane
induces a ``spin-flop'' transition, where the noncollinear AF spin
structure of PLCCO is transformed into a collinear one. The
spin-flop transition is continuous in semiconducting PLCCO, but
gradually becomes sharp with increasing doping and the appearance
of superconductivity. A $c$-axis aligned magnetic field that
suppresses the superconductivity also enhances the quasi-2D SDW
order at $(0.5,0.5,0)$ for underdoped PLCCO. However, there is no
effect on the 3D AF order in either superconducting or
nonsuperconducting samples. Since the same field along the
$[\bar{1},1,0]$ direction in the CuO$_2$ plane has no (or little)
effect on the superconductivity, $(0.5,0.5,0)$ and
(Pr,La,Ce)$_2$O$_3$ impurity positions, we conclude that the
$c$-axis field-induced effect is intrinsic to PLCCO and arises
from the suppression of superconductivity.

\end{abstract}

\pacs{74.72.Jt, 75.25.+z, 75.50.Ee, 61.12.Ld}

\maketitle \narrowtext
\section{Introduction}
All high-transition-temperature (high-$T_c$) copper oxides
(cuprates) have insulating antiferromagnetic (AF), superconducting
(SC), and metallic phases \cite{kastner}. Understanding the nature
of these coexisting and competing phases is a key challenge of
current research in high-$T_c$ superconductivity
\cite{sczhang,levi}. While the SC phase in cuprates arises from
hole ($p$-type) or electron ($n$-type) doping of the insulating AF
phase, it is still unclear how the superconductivity is related to
the AF phase \cite{sczhang}.

In the hole-doped La$_{2-x}$Sr$_x$CuO$_4$ (LSCO) near $x\sim 1/8$
\cite{Kimura} and La$_2$CuO$_{4+y}$ \cite{Lee}, neutron scattering
experiments have revealed the presence of a static incommensurate
spin-density-wave (SDW) order in coexistence with the
superconductivity. The SDW is quasi-two-dimensional (2D) with
long-range order in the CuO$_2$ plane and short-range correlations
along the $c$-axis \cite{Kimura,Lee}. Indexed on a tetragonal unit
cell of LSCO, where the AF order occurs at the reciprocal lattice
position $(0.5,0.5)$ in the CuO$_2$ plane, the positions of the
four incommensurate SDW peaks are $(0.5\pm\Delta H,0.5)$ and
$(0.5,0.5\pm\Delta K)$ as shown in Fig. 1b \cite{kastner}. At
optimal doping (highest $T_c$), the static SDW order in LSCO
vanishes and is replaced by the appearance of a spin gap and
incommensurate spin fluctuations at energies above the gap
\cite{kyamada}. On the other hand, experiments on lightly doped
insulating ($x=0.03$, 0.04, 0.05) and underdoped SC ($x=0.06$)
LSCO have shown that the static SDW order appears in these
samples, but intriguingly the four incommensurate peaks in Fig. 1b
are only observed with the establishment of the bulk
superconductivity for the underdoped SC samples ($0.06\leq x \leq
0.12$) \cite{wakimoto,fujitalsco,matsudalsco}. Therefore, the
insulating-to-superconducting phase transition induced by
increasing hole(Sr)-doping is associated with the appearance of a
static quasi-2D SDW modulation.

\begin{figure}
\includegraphics[keepaspectratio=true,width=0.8\columnwidth,clip]{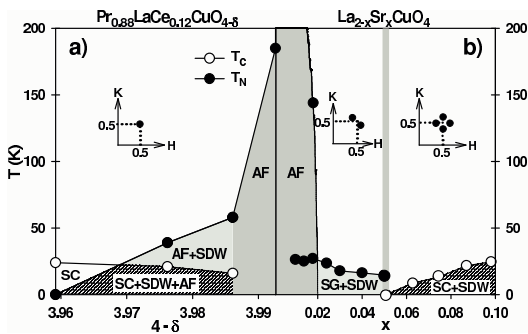}
\caption{Comparison of the phase diagrams between hole-doped
La$_{2-x}$Sr$_x$CuO$_4$ (LSCO) and electron-doped
Pr$_{0.88}$LaCe$_{0.12}$CuO$_{4\pm\delta}$ (PLCCO). a) Properties
of the PLCCO samples investigated in this work. Since the exact
oxygen concentrations in our samples are unknown, the values of
$4-\delta$ in the horizontal axis are obtained using Fig. 8 of
Ref. \cite{jskim} by assuming that PLCCO samples with the same
$T_c$'s as those of Nd$_{1.85}$Ce$_{0.15}$CuO$_{4-\delta}$ have
the same oxygen content. When superconductivity is first
established in PLCCO, a commensurate static SDW order also
appears. The open and filled circles are data from Ref.
\cite{daiprl}. b) Phase diagram of LSCO as a function Sr-doping
from neutron scattering results of Ref.
\cite{wakimoto,fujitalsco,matsudalsco}. In the spin-glass (SG)
phase of LSCO, two static incommensurate SDW peaks at $(0.5+\Delta
H,0.5-\Delta K)$ and $(0.5-\Delta H,0.5+\Delta K)$ (see inset) are
observed for $0.02\leq x\leq 0.05$. When superconductivity sets in
at $x\geq 0.05$, two pairs of incommensurate SDW peaks appear
simultaneously at $(0.5\pm\Delta H,0.5)$ and $(0.5,0.5\pm\Delta
K)$. The data in b) are from Refs.
\cite{wakimoto,fujitalsco,matsudalsco}.}
\end{figure}

When a magnetic field is applied along the $c$-axis of the
underdoped SC LSCO, it not only suppresses the superconductivity
but also enhances the zero-field SDW order
\cite{Katano,Lake1,Khaykovich,Khaykovich1}. For optimally doped
LSCO without static SDW order, a $c$-axis magnetic field induces
spin fluctuations below the spin-gap energy and drives the system
toward static AF order \cite{lake2,tranquada}. Since the same
field applied in the CuO$_2$ plane has little effect on
superconductivity or the SDW order \cite{Lake}, these results
suggest that antiferromagnetism is a competing ground state
revealed by the suppression of superconductivity
\cite{Zhang,Arovas,demler,dhlee,ychen,skivelson,jzhu}.

If suppression of superconductivity in hole-doped LSCO indeed
drives the system toward an AF ordered state, it is important to
determine the universality of such a feature in other SC copper
oxides \cite{dhz}. Although neutron scattering experiments failed
to confirm any enhancement of the static long-range AF order in
hole-doped YBa$_2$Cu$_3$O$_{6.6}$ for a $c$-axis aligned field up
to 7-T \cite{dainature,mookprb}, these measurements may not be
conclusive because of the enormous upper critical fields $B_{c2}$
($>45$-T for $c$-axis aligned fields) required to completely
suppress superconductivity. Compared to hole-doped cuprates,
electron-doped materials offer a unique opportunity for studying
the magnetic field-induced effect for two reasons. First,
electron-doped materials generally have upper critical fields less
than 15-T \cite{hidaka,fournier,wang2003}, a value reachable in
neutron scattering experiments. Second, electron-doped materials
can be transformed from as-grown nonsuperconducting (NSC) AF
insulators to full superconductivity by simply annealing the
samples at different temperatures to modify the charge carrier
density \cite{tokura,takagi,jskim,kurahashi,fujita1}. This unique
property allows the possibility of studying the NSC to SC
transition in electron-doped cuprates without the complications of
structural and/or chemical disorder induced by chemical
substitution in hole-doped materials. Recently, we discovered that
when superconductivity first appears in one family of
electron-doped superconductors,
Pr$_{0.88}$LaCe$_{0.12}$CuO$_{4\pm\delta}$ (PLCCO), a commensurate
quasi-2D SDW order is also induced \cite{notesdw}, and both
coexist with the residual three-dimensional (3D) AF state
\cite{daiprl}. The optimally doped SC PLCCO, on the other hand,
has no static SDW or residual AF order \cite{fujita1,daiprl},
analogous to hole-doped LSCO as depicted in Fig. 1
\cite{kyamada,wakimoto,fujitalsco,matsudalsco}.

\begin{figure}
\includegraphics[keepaspectratio=true,width=0.8\columnwidth,clip]{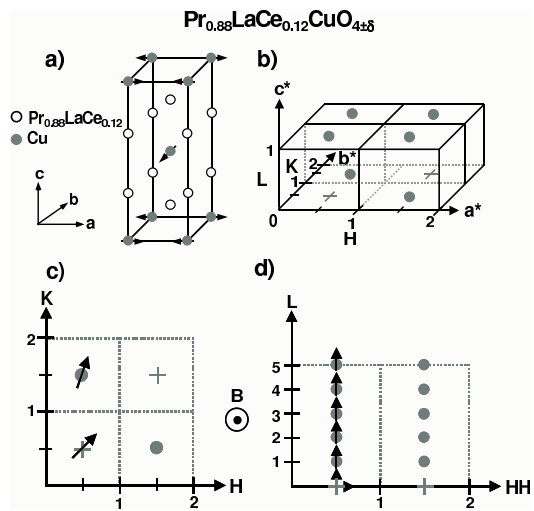}
\caption{a) Schematic diagram of the noncollinear spin structure
of PLCCO, b) Allowed AF Bragg Peaks from the spin structure of a),
c) Reciprocal space probed in the $[H,K,0]$ scattering plane where
the applied field is along the $c$-axis. Solid circles are allowed
AF Bragg peak positions in the $[H,K,0]$ plane. d) Similar
experiments in the $[H,H,L]$ zone where the field is along the
$[\bar{1},1,0]$ direction in the CuO$_2$ plane. Arrows indicate
the scan directions in our experiments.}
\end{figure}

For prototypical electron-doped Nd$_{2-x}$Ce$_x$CuO$_4$ (NCCO)
\cite{tokura,takagi}, previous neutron scattering experiments show
a drastic suppression of the static 3D AF order when
superconductivity is established \cite{yamada}. However, the
static AF order persists even for NCCO with optimal bulk
superconductivity ($T_c=25$ K) \cite{yamadaprl}. In the initial
neutron scattering experiments on the slightly underdoped SC
Nd$_{1.86}$Ce$_{0.14}$CuO$_4$, Matsuda {\it et al.} \cite{matsuda}
found that a 10-T $c$-axis-aligned magnetic field has no effect on
the residual AF order in the sample. On the other hand, we
discovered that such fields enhance magnetic intensities at AF
ordering positions in an optimally doped
Nd$_{1.85}$Ce$_{0.15}$CuO$_4$ \cite{Kang,Matsuura}. The induced AF
moments scale approximately linearly with the applied field,
saturate at $B_{c2}$, and then decrease for higher fields. These
results thus indicate that AF order is a competing ground state to
superconductivity in electron-doped Nd$_{1.85}$Ce$_{0.15}$CuO$_4$
\cite{Chen}.

Although electron-doped NCCO offers a unique opportunity for
studying the superconductivity-suppressed ground state of
high-$T_c$ cuprates, the system has one important complication.
The as-grown material is NSC, and has to be oxygen reduced to
render the system superconducting. This reducing process has been
found to produce a small quantity of the cubic (Nd,Ce)$_2$O$_3$ as
an impurity phase \cite{Matsuura,Mang,Kang1,Matsuura1,Mang1}.
Generally, a small amount (0.01\% to 1\%) of randomly distributed
impurity would be unobservable, but (Nd,Ce)$_2$O$_3$ stabilizes as
an oriented crystalline lattice in the NCCO matrix because its
lattice constant is about 2$\sqrt{2}$ larger than the planner
lattice constant of the tetragonal NCCO ($a=3.945$ \AA\ and
$a_{NO}=2\sqrt{2}a$). This in-plane lattice match with NCCO means
reflections at $(H, K, 0)$ of NCCO in reciprocal space may also
stem from (Nd,Ce)$_2$O$_3$, thus giving structural impurity peaks
that match some of the magnetic peaks [such as $(0.5,0.5,0)$]. In
the paramagnetic state of (Nd,Ce)$_2$O$_3$, a field induces a net
magnetization on Nd$^{3+}$, which can then contaminate the
intensity of the cuprate magnetic peak such as $(0.5,0.5,0)$
\cite{Matsuura,Mang,Kang1,Matsuura1,Mang1}.

There are three ways to resolve this impurity problem and
determine the intrinsic properties of electron-doped materials.
First, since the impurity and cuprate peaks are not lattice
matched along the $c$-axis direction (the lattice constant of
impurity is about 10\% smaller than that of NCCO), the impurity
peaks occur at positions such as $(0.5, 0.5, L)$ $L=2.2$, 4.4, and
so on \cite{Matsuura,Mang}. This allows the impurity scattering to
be determined separately, and the in-plane scattering can then be
corrected by subtracting the impurity contribution to ascertain
the NCCO contribution of the in-plane peaks
\cite{Matsuura,Matsuura1}. Second, with a $c$-axis-aligned
magnetic field and $c$-axis in the scattering plane, the non-zero
integer $L$ positions ($L = 1$, 2, 3, $\cdots$) can be measured
without any possible impurity contribution. We carried out such an
experiment using a horizontal field magnet and found that
application of a 4-T $c$-axis aligned magnetic field enhances the
$(0.5, 0.5, 3)$ AF reflection in the optimally doped SC
Nd$_{1.85}$Ce$_{0.15}$CuO$_4$ \cite{Matsuura}. Both procedures
show unambiguously that there is a field-induced moment in
electron-doped Nd$_{1.85}$Ce$_{0.15}$CuO$_4$, whose behavior is
similar to that observed in the hole-doped systems.

\begin{figure}
\includegraphics[keepaspectratio=true,width=0.8\columnwidth,clip]{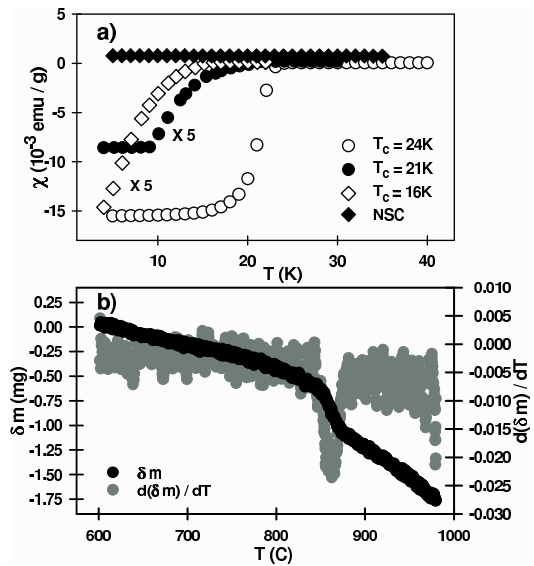}
\caption{a) Magnetic susceptibility measurements of
Pr$_{0.88}$LaCe$_{0.12}$CuO$_{4\pm\delta}$ single crystals under a
small (5-10 Gauss) field in the CuO$_2$ plane. b) Temperature
dependence of weight change and its derivative of a
Pr$_{1.18}$La$_{0.7}$Ce$_{0.12}$CuO$_{4-\delta}$ in the annealing
process.}
\end{figure}

Although our previous experiments \cite{Kang,Matsuura,Matsuura1}
have demonstrated the intrinsic nature of the field-induced effect
in SC Nd$_{1.85}$Ce$_{0.15}$CuO$_4$, it is still interesting to
perform field-induced experiments on samples free of Nd because
application of a magnetic field is known to induce a large moment
on rare-earth Nd$^{3+}$ spins that may conceal the intrinsic field
dependence of the Cu$^{2+}$ moments. Electron-doped
Pr$_{1-x}$LaCe$_{x}$CuO$_{4\pm\delta}$, where Pr$^{3+}$ has a
singlet ground state \cite{boothroyd} and La is nonmagnetic, is
ideal for this purpose. Using single crystals of
Pr$_{1-x}$LaCe$_{x}$CuO$_{4\pm\delta}$ with $x=0.11$ ($T_c=16$ K)
and $x=0.15$ ($T_c=26$ K), Fujita {\it et al.} \cite{fujitaprl}
found that a $c$-axis aligned magnetic field enhances the residual
AF order in the $x=0.11$ sample at $(0.5,1.5,0)$ position, but has
no effect in the overdoped $x=0.15$ sample. The authors conclude
that a $c$-axis field changes both the magnetic intensity and the
onset of the N\'{e}el ordering temperature $T_N$ for the $x=0.11$
sample, with the maximum field effect at $\sim$5-T
\cite{fujitaprl}.

Our approach to this problem has been slightly different from that
of Fujita {\it et al.} \cite{fujitaprl}. Instead of preparing
different SC Pr$_{1-x}$LaCe$_{x}$CuO$_{4\pm\delta}$ samples as a
function of Ce-doping, we grew single crystals of PLCCO using a
traveling solvent floating zone furnace. As-grown, PLCCO has a
noncollinear Cu spin structure as shown in Fig. 2a with a small
induced Pr moment due to Cu-Pr interaction \cite{Lavrov}. When the
AF ordered semiconducting PLCCO is transformed into a
superconductor by annealing in pure Ar, the SC transition
temperatures ($T_c$'s) can be controlled by judicially tuning the
annealing temperature \cite{fujita1}. This approach avoids
complications of Ce-substitution for different $T_c$ materials.
For underdoped samples with $T_c$'s below 25 K, superconductivity
coexists with both the 3D AF-type order of the undoped material
and a commensurate quasi-2D SDW modulation \cite{daiprl}.
Optimally doped PLCCO has no static SDW or residual AF order.

In this article, we present comprehensive neutron scattering
studies of four single crystals of $n$-type PLCCO from a
semiconductor with $T_N=186$ K to superconductors with $T_c$ = 16
K, 21 K, and 24 K, respectively (Figs. 1a and 3a). When a magnetic
field is applied along the $[\bar{1},1,0]$ direction parallel to
the CuO$_2$ plane, the noncollinear AF spin structure of PLCCO is
transformed into a collinear one through a ``spin-flop''
transition \cite{Lavrov}. Such a spin-flop transition is
continuous in semiconducting PLCCO but gradually changes to
discontinuous with increasing doping and the appearance of
superconductivity. For SC PLCCO, an in-plane field only induces a
spin-flop transition but does not change the effective Cu$^{2+}$
or Pr$^{3+}$ moments for fields up to 14-T. A $c$-axis aligned
magnetic field, on the other hand, enhances the quasi-2D SDW order
at $(0.5,0.5,0)$ for underdoped PLCCO, but has no effect on the 3D
AF order in the NSC and SC samples. Since the same 14-T field
along the $[\bar{1},1,0]$ direction parallel to the CuO$_2$ plane
has no effect on $(0.5,0.5,0)$ and on the lattice Bragg peaks of
the (Pr,La,Ce)$_2$O$_3$ impurity phase, we conclude that the
$c$-axis field-induced effect is intrinsic to PLCCO and arises
from the suppression of superconductivity.

The rest of the paper is organized as follows: in Sec. II we
describe the details of sample preparation methods and
neutron-scattering experimental setup. The main experimental
results are summarized in Sec. III, where Sec. III.A reviews the
zero-field magnetic properties, Sec. III.B discusses the effects
of in-plane magnetic fields on the AF spin structure, Sec. III.C
summarizes the magnetic field effect on the NSC PLCCO and impurity
(Pr,La,Ce)$_2$O$_3$ phase, Sec. III.D presents the comprehensive
$c$-axis field effect studies and their temperature dependence for
different SC PLCCO samples, and finally Sec. III.E discuss the
anisotropy of the field effect by comparing the in-plane and
$c$-axis field effect on the $T_c=21$ K SC sample. In Sec. IV, we
compare our results with previous experiments in $p$- and $n$-type
materials. A brief summary is given in Sec. V.

\section{Experimental details}

We grew high quality PLCCO single crystals (cylindrical rods
weight 0.8-1.5 grams with mosaicity $<1^\circ$) using the
traveling solvent floating zone technique and annealed the samples
in pure Ar at different temperatures to control the SC transition
temperatures. The partial substitution of Pr with La was used to
stabilize the crystal growth without introducing significant
lattice distortions \cite{Sun}. We obtained three SC PLCCO samples
with the onset temperature for bulk superconductivity at $T_c =
24$ K, 21 K, and 16 K from magnetic susceptibility measurements
(Fig. 3a). The $T_c = 24$ K, and 21 K samples are obtained by
annealing the as-grown single crystals in pure argon at
$970^\circ$C and $940^\circ$C for 24 hours, respectively. To
obtain the $T_c = 16$ K crystal, the as-grown sample was annealed
in pure argon at $915^\circ$C for 1 week. The SC PLCCO of $T_c =
21$ K and 16 K are near the phase boundary between AF and SC
phases and the $T_c = 24$ K sample is in the optimally doped
regime. We also obtained a NSC sample by annealing one of the SC
$T_c = 16$ K samples in air at $900^\circ$C for 24 hours
\cite{daiprl}.

One of the open questions in electron-doped superconductors is the
role of the annealing process to superconductivity \cite{tokura}.
In earlier works, it was postulated that as-grown NSC samples have
excess oxygen atoms above copper sites (apical oxygen). These
extraneous oxygen atoms have been believed to induce a local
disordered potential that localizes doped electrons and therefore
prohibits superconductivity \cite{Jiang}. When the annealing
process removes these defect oxygen atoms at apical sites, the
electrons will not be localized and the superconductivity appears.
The magnitude of the Hall coefficient in annealed NCCO ($x =
0.15$) decreases dramatically below $T<100$ K, very differently
from the as-grown sample \cite{Jiang}. This suggests that
annealing changes the mobile charge density. To understand what
annealing does to our PLCCO samples, we have performed
thermogravimetric analysis (TGA) of our samples under different
annealing conditions. Figure 3b shows the temperature dependence
of the weight change for a $\rm
Pr_{1.18}La_{0.7}Ce_{0.12}CuO_{4\pm\delta}$ sample during the
annealing process. The large drop in weight change shows up as a
kink in the data and a sharp dip in its derivative; we interpret
this kink to be indicative of the point where the oxygen content
becomes the closest to stoichiometry, namely $\delta\simeq 0$
\cite{Sun}. In any case, the TGA data give the most direct
evidence that oxygen content of the sample is reduced during the
annealing process.

\begin{figure}
\includegraphics[keepaspectratio=true,width=0.8\columnwidth,clip]{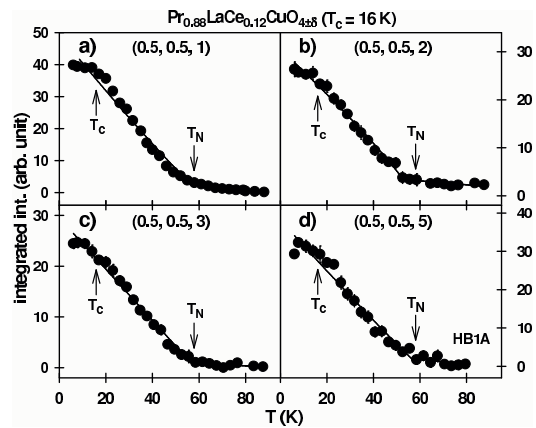}
\caption{Temperature dependence of integrated intensities at
$(0.5, 0.5, L)$ for the $T_c=16$ K PLCCO. The solid lines are
guides to the eye. The arrows indicate $T_c$ and $T_N$,
respectively.}
\end{figure}

While TGA analysis indicates a decreased oxygen content after the
reduction process, it is still unclear whether the apical oxygen
is removed \cite{Radaelli,Nath}. Indeed, recent Raman and
crystal-field infrared transmission results suggest that instead
of removing apical oxygen as originally thought, reduction of
optimally doped Pr$_{2-x}$Ce$_x$CuO$_{4\pm\delta}$ and NCCO
samples actually involves only oxygens in the CuO$_2$ plane
\cite{Riou,Richard}. Clearly, the exact oxygen content and their
arrangements in the crystal before and after the reduction need to
be sorted out by future studies.

Another issue to be noted regarding the PLCCO material is the
relevance of the (Pr,La,Ce)$_2$O$_3$ impurity phase, which plays
the same role in PLCCO as the (Nd,Ce)$_2$O$_3$ phase in NCCO. To
understand the effect of annealing on this impurity phase, we have
performed systematic synchrotron X-ray diffraction investigations
on as-grown NSC, SC, and re-oxygenated NSC PLCCO samples at the
Advanced Photon Source, Argonne National Laboratory. We confirm
that the (Pr,La,Ce)$_2$O$_3$ impurity phase can be produced by
annealing and but it vanishes after the reoxygenation process
\cite{kurahashi}; such a behavior is rather strange, but is
understandable if the host PLCCO crystals are slightly Cu
deficient. To confirm this possibility, inductively-coupled plasma
atomic-emission spectroscopy (ICP-AES) analysis was performed on a
batch of crystals that are nominally
Pr$_{1.18}$La$_{0.7}$Ce$_{0.12}$CuO$_{4\pm\delta}$. The reference
solution is prepared by diluting commercial standard solutions not
by the volume but by the weight, which enables us to calculate the
cation concentration of the reference with a high resolution. A
piece of the Pr$_{1.18}$La$_{0.7}$Ce$_{0.12}$CuO$_{4\pm\delta}$
crystal (1 mg) is dissolved into a 30 cm$^3$ of HNO$_3$ (1
Mol/Liter). The analyzed cation ratio of the nominally
Pr$_{1.18}$La$_{0.7}$Ce$_{0.12}$CuO$_{4\pm\delta}$ sample was
${\rm Pr : La : Ce : Cu = 1.20 : 0.68 : 0.12 : 0.97}$, with a
relative error of less than 1\% for each element. This result
shows that Cu is indeed deficient by about 3\% in the
Pr$_{1.2}$La$_{0.7}$Ce$_{0.12}$CuO$_{4\pm\delta}$ crystal, which
is probably also true for PLCCO. Assuming that as-grown PLCCO
samples are slightly Cu deficient as
Pr$_{1.2}$La$_{0.7}$Ce$_{0.12}$CuO$_{4\pm\delta}$, the effect of
annealing is then to remove oxygen in PLCCO to form the
(Pr,La,Ce)$_2$O$_3$ impurity and make the remaining crystal
structure more perfect for superconductivity. This analysis can
potentially explain why the (Pr,La,Ce)$_2$O$_3$ impurity phase can
be reversibly produced \cite{kurahashi}.

Our neutron scattering measurements were performed on the HB-1A,
HB-1, and HB-3 triple-axis spectrometers at the high-flux-isotope
reactor (HFIR), Oak Ridge National Laboratory (ORNL) and on the E4
two-axis diffractometer and E1 triple-axis spectrometer at the
Berlin Neutron Scattering Center, Hahn-Meitner-Institute (HMI).
The field effect experiments on SC samples of $T_c = 24$ K, 21 K,
and 16 K and the NSC sample ($T_N=186$ K) were carried out at HFIR
using a 7-T vertical field SC magnet \cite{dainature}. The
collimations were
48$^\prime$-40$^\prime$-sample-40$^\prime$-102$^\prime$
(full-width at half maximum) from the reactor to the detector and
the final neutron energies were fixed at either $E_f=14.6$ meV or
13.5 meV. For high magnetic field experiments, we used the VM-1
14.5-T vertical field SC magnet at HMI and a
40$^\prime$-40$^\prime$-sample-40$^\prime$ collimation with a
fixed neutron final energy of 13.6 meV. A pyrolytic graphite (PG)
monochromator and PG analyzer were used and PG filters were placed
in front of the sample to remove higher order contaminations from
the incident beam. The samples were clamped on solid aluminum
brackets similar to those used in the NCCO experiments
\cite{Matsuura}. However, a $c$-axis aligned field almost exerts
no torque on PLCCO because of the nonmagnetic singlet nature of
the Pr ground state.

\begin{figure}
\includegraphics[keepaspectratio=true,width=0.8\columnwidth,clip]{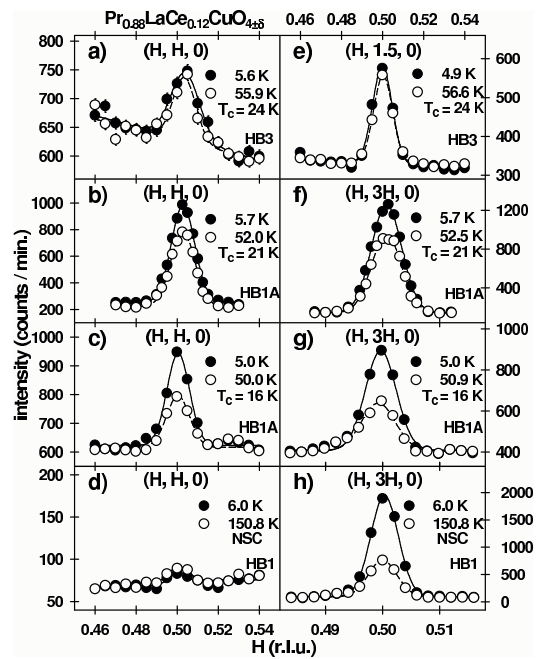}
\caption{\textbf{Q} scans in the $[H,K,0]$ zone at low and high
temperatures for different PLCCO samples. a-d) Scans along the
$[H,H,0]$ direction around $(0.5,0.5,0)$ for $T_c=24$ K, 21 K, 16
K, and NSC PLCCO, e-h) Similar scans along the $[H,3H,0]$
direction around $(0.5,1.5,0)$. This figure is reproduced from
Fig. 2 of Ref. \cite{daiprl}. }
\end{figure}

We use elastic neutron scattering to probe AF order and label the
momentum transfer ${\bf Q}=(q_x,q_y,q_z)$ in \AA$^{-1}$ as
$(H,K,L)=(q_xa/2\pi,q_yb/2\pi,q_zc/2\pi)$ in the reciprocal
lattice units (r.l.u.) appropriate for the tetragonal unit cell of
PLCCO (space group $I4/mmm$, $a=b=3.98$ and $c = 12.27$ \AA). Here
$a$ and $c$ are in-plane and out-of -plane lattice constants,
respectively. The crystals were aligned in two different
geometries, $[H,K,0]$ and $[H,H,L]$ scattering planes. The
vertical magnetic field was applied along the $c$-axis in the
former scattering plane and was applied along the $[\bar{1},1,0]$
direction (${\bf B}$$\,\parallel\,$$ab$-plane) in the later case.
The $[H,H,L]$ geometry was also used to search for quasi-2D SDW
order at zero-field using a two-axis energy integrated mode (Fig.
2d), to determine the field-induced effect on the impurity phase,
and investigate the field-induced spin-flop transitions. In the
$[H,K,0]$ geometry, we studied the $c$-axis field-induced effect
on SC and NSC PLCCO. The anisotropy of the field-induced effect
can be determined by comparing the results in these two geometries
since the magnetic field along the $c$-axis suppresses the SC much
more strongly than that for the same field in the $ab$-plane.

\section{Results}

\subsection{Magnetic properties of superconducting (SC) and nonsuperconducting (NSC)
Pr$_{0.88}$LaCe$_{0.12}$CuO$_{4\pm \delta}$ at zero-field}

To determine the influence of a magnetic field on the AF order of
PLCCO, it is helpful to know the magnetic structure of its parent
compound. Pr$_2$CuO$_4$ has a body-centered tetragonal structure
with space group $I4/mmm$ \cite{sumarlin}. The spins of Cu$^{2+}$
ions have AF ordering in the CuO$_2$ plane due to strong in-plane
exchange interaction. The orientation of spins in adjacent planes
is noncollinear governed by the weak pseudo-dipolar interaction
between planes, since exchange field on each copper ion by
neighboring planes is canceled due to body-centered tetragonal
crystal symmetry \cite{sumarlin}. The spins at $(0, 0, 0)$ and
$(0.5, 0.5, 0.5)$ positions are along the $[1,0,0]$ and
$[0,\bar{1},0]$ directions, respectively (Fig. 2a). In this
magnetic structure, the AF Bragg peak at $(0.5, 0.5, 0)$ is
disallowed. The allowed magnetic Bragg peaks are at $(0.5,0.5,L)$
($L=1,2,3,\cdots$) and $(0.5,1.5,L)$ ($L=0,1,2,3,\cdots$) marked
as solid dots in Fig. 2b.

\begin{figure}
\includegraphics[keepaspectratio=true,width=0.8\columnwidth,clip]{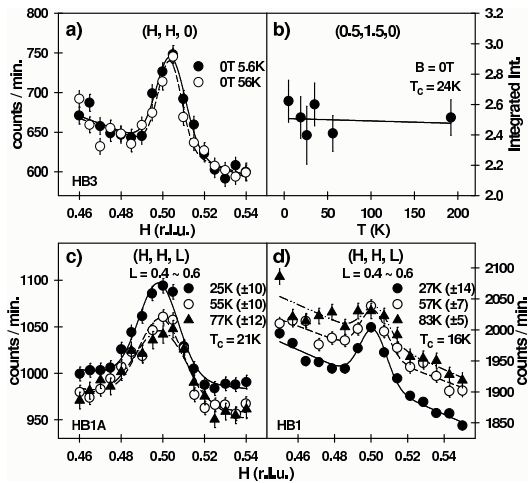}
\caption{a) $[H,H,0]$ scans around $(0.5, 0.5, 0)$ at 5.6 K and 56
K for the optimally doped $T_c=24$ K PLCCO. The observed peak at
$(0.5,0.5,0)$ has nonmagnetic origin. b) The temperature
dependence of the integrated intensities at $(0.5,1.5,0)$ for the
$T_c=24$ K PLCCO. There is no evidence for static SDW and/or AF
orders in the optimally doped PLCCO. c-d) Two-axis mode scans
along the $[H,H,0]$ direction around $(0.5, 0.5, L)$ ($0.4<L<0.6$)
at different temperatures for underdoped SC $T_c=21$ K and 16 K
PLCCO, respectively. Magnetic scattering clearly decreases with
increasing temperature for both compounds.}
\end{figure}

We performed neutron diffraction measurements at zero-field on
magnetic Bragg peaks $(0.5, 0.5, L)$ ($L=1,2,3,5$) to determine
the magnetic structure of the superconducting $T_c = 16$ K PLCCO
(Fig. 4). In principle, one cannot uniquely determine the spin
structure of PLCCO based on magnetic intensities at $(0.5, 0.5,
L)$ alone. However, since the spin structure of
Pr$_{2-x}$Ce$_x$CuO$_4$ at all Ce doping levels is non-collinear
as depicted in Fig. 2a \cite{sumarlin}, one would expect a similar
magnetic structure in PLCCO because substitution of La$^{3+}$ for
Pr$^{3+}$ will weaken the pseudo-dipolar interactions but not
eliminate it. The integrated intensity comparison of $(0.5, 0.5,
L)$ peaks to the calculated intensities confirms that the
superconducting sample has the same noncollinear spin structure as
that of the parent Pr$_2$CuO$_4$.

To estimate the effective Cu$^{2+}$ and Pr$^{3+}$ moments, we
measured the temperature dependence of $(0.5, 0.5, L)$
reflections. By comparing the magnetic structure factor
calculations (Table I) with that of the weak nuclear Bragg peak
$(1,1,0)$, we calculate an ordered Cu$^{2+}$ moment of
$0.06\pm0.01$ $\mu_B$ at 5 K with negligible induced moment on
Pr$^{3+}$ ions for the $T_c = 16$ K PLCCO \cite{note1}. Since the
induced Pr$^{3+}$ moment contributes positively to the intensity
of $(0.5, 0.5,1)$ and $(0.5, 0.5,2)$, but negatively to $(0.5,
0.5, 3)$, the temperature dependent measurements of $(0.5, 0.5,
L)$ should reveal deviations from power law fits when the
Pr-induced moment becomes significant \cite{Lavrov}. In addition,
the $(0.5, 0.5, 5)$ peak should exhibit only the temperature
dependence of the Cu$^{2+}$ moment as there are essentially no
Pr$^{3+}$ moment contributions to this reflection (Table I). Since
all magnetic peaks at $(0.5, 0.5, L)$ show a similar temperature
dependence as $(0.5, 0.5, 5)$ (Fig. 4), we conclude that there is
negligible Pr-induced moment above 4 K in the $T_c = 16$ K PLCCO.

\begin{figure}
\includegraphics[keepaspectratio=true,width=0.8\columnwidth,clip]{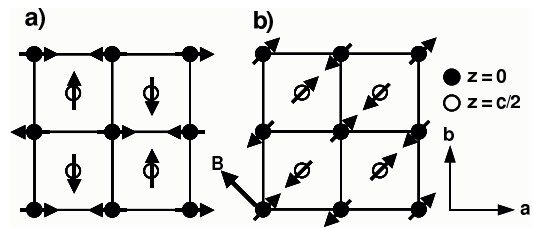}
\caption{Magnetic structures of Cu$^{2+}$ ions in PLCCO. a)
Noncollinear structure at zero-field. b) Collinear structure
transformed by a magnetic field applied along the $[\bar{1},1,0]$
direction.}
\end{figure}

In a recent work, we have systematically investigated how
superconductivity develops from the AF ordered insulating state in
PLCCO \cite{daiprl}. By carefully tuning the annealing temperature
of as-grown NSC PLCCO, we can control the strength of the AF order
and superconductivity without additional complications such as
disorder associated with conventional chemical substitution. The
temperature dependence of $(0.5, 1.5, 0)$ shows that the N\'{e}el
temperature $T_N$ decreases linearly with increasing $T_c$, and
becomes zero at the maximum $T_c$ of 24 K. Although the
commensurate AF position $(0.5,0.5,0)$ is forbidden in the
noncollinear spin structure of PLCCO (Fig. 2a), clear magnetic
scattering is observed at $(0.5,0.5,0)$ in underdoped SC PLCCO
($T_c = 21$ K and 16 K) in the $[H,K,0]$ scattering geometry (Fig.
5). This magnetic component is on top of the structural
superlattice peak resulting from the annealing process
\cite{kurahashi}. Solid lines in Fig. 5 are Gaussian fits to the
peaks on sloping backgrounds, where $I=bkgd+I_0
\exp[-(H-H_0)^2/(2\sigma^2)]$. Fourier transform of the Gaussian
peaks give minimum a in-plane coherence lengths ($CL$) of
$\sim$200 \AA\ for all four samples using
$CL=[\sqrt{\ln(2)}/\pi](a/\sigma)$ for $[H,H,0]$ scans and
$CL=[\sqrt{\ln(2)/5}/\pi](a/\sigma)$ for $[H,3H,0]$ scans. To
determine the $c$-axis correlations of the magnetic scattering at
$(0.5,0.5,0)$, we aligned the crystal in the $[H,H,L]$ scattering
plane but found no large intensity gain centered at $L=0$ in
$[0.5,0.5,L]$ scans \cite{daiprl}.

If underdoped PLCCO has a quasi-2D SDW with strong correlation in
the CuO$_2$ plane but weak coupling along the \textit{c}-axis, the
coarse vertical resolution of the triple-axis spectrometer will
integrate a much larger region of the $c$-axis magnetic rod in the
$[H,K,0]$ plane than that in the $[H,H,L]$ zone. As a consequence,
the quasi-2D scattering should be more easily observable in the
$[H,K,0]$ zone. The presence of a quasi-2D SDW can be tested using
the 2-axis energy integrated mode by aligning the outgoing wave
vector $k_f$ parallel to the 2D rod direction ($c$-axis) (Fig. 4d
in Ref. \cite{daiprl}). The temperature dependent scattering
around $(0.5,0.5,L)$ in underdoped PLCCO ($T_c = 21$ and 16 K)
suggests the presence of quasi-2D SDW modulation (Figs. 6c and 6d)
\cite{daiprl}. Because these data were collected using the 2-axis
mode, which integrates elastic, inelastic magnetic scattering as
well as phonons, its temperature dependence is not a reliable
measure of the quasi-elastic diffusive magnetic scattering in Fig.
3 of Ref. \cite{daiprl}. For optimally doped PLCCO ($T_c = 24$ K),
there is no evidence for the SDW or 3D AF order as shown by the
temperature independent scattering at $(0.5, 0.5, 0)$ and $(0.5,
1.5, 0)$ (Figs. 5a, 5e, 6a, and 6b). For NSC PLCCO, the scattering
only shows the static 3D AF order with a small Pr-induced moment
at low temperatures \cite{daiprl}.

\begin{table}
\caption{Magnetic structure factor calculations for noncollinear
$(F_{nc})^2$ and collinear $(F_c)^2$ spin structures as seen in
Figures 7a and 7b. Here $\gamma e^2/(2mc^2)=0.2695\times 10^{-12}$
cm, $f_{cu,pr}$, $M_{cu,pr}$ are magnetic form factors and ordered
magnetic moments for Cu$^{2+}$ and Pr$^{3+}$ ions, respectively. }
\begin{ruledtabular}
\begin{tabular}{cc}
$(H,K,L)$ & $(F_{nc})^2$ \\
\hline
$(0.5,0.5,0)$ & 0 \\
$(0.5,0.5,1)$ & 32.0 ($\gamma$e$^2$/2$m$c$^2$)$^2$ ($f_{cu}M_{cu}$ + 0.5310$f_{pr}M_{pr})^2$ \\
$(0.5,0.5,2)$ & 14.6 ($\gamma$e$^2$/2$m$c$^2$)$^2$ ($f_{cu}M_{cu}$ + 0.2391$f_{pr}M_{pr})^2$ \\
$(0.5,0.5,3)$ & 32.0 ($\gamma$e$^2$/2$m$c$^2$)$^2$ ($f_{cu}M_{cu}$ $-$ 0.8196$f_{pr}M_{pr})^2$ \\
$(0.5,0.5,4)$ & 24.6 ($\gamma$e$^2$/2$m$c$^2$)$^2$ ($f_{cu}M_{cu}$ + 0.7500$f_{pr}M_{pr})^2$ \\
$(0.5,0.5,5)$ & 32.0 ($\gamma$e$^2$/2$m$c$^2$)$^2$ ($f_{cu}M_{cu}$ $-$ 0.0856$f_{pr}M_{pr})^2$ \\
\end{tabular}
\end{ruledtabular}

\vspace{3 mm}

\begin{ruledtabular}
\begin{tabular}{cc}
$(H,K,L)$ & $(F_c)^2$ \\
\hline
$(0.5,0.5,0)$ & 0 \\
$(0.5,0.5,1)$ & 0 \\
$(0.5,0.5,2)$ & 29.2 ($\gamma$e$^2$/2$m$c$^2$)$^2$ ($f_{cu}M_{cu}$ + 0.2391$f_{pr}M_{pr})^2$ \\
$(0.5,0.5,3)$ & 0 \\
$(0.5,0.5,4)$ & 49.2 ($\gamma$e$^2$/2$m$c$^2$)$^2$ ($f_{cu}M_{cu}$ + 0.7500$f_{pr}M_{pr})^2$ \\
$(0.5,0.5,5)$ & 0 \\
\end{tabular}
\end{ruledtabular}
\end{table}

\subsection{Spin-flop transition on superconducting (SC) and nonsuperconducting (NSC)
Pr$_{0.88}$LaCe$_{0.12}$CuO$_{4\pm\delta}$}

A magnetic field applied in the $[\bar{1}, 1, 0]$ direction of the
CuO$_2$ planes induces a spin-flop transition from noncollinear to
collinear spin structure in electron-doped parent compounds
Nd$_2$CuO$_4$ and Pr$_2$CuO$_4$ (Fig. 7) \cite{Skanthakumar,
Plakhty}. For lightly electron-doped $\rm
Pr_{1.29}La_{0.7}Ce_{0.01}CuO_{4\pm\delta}$, Lavrov {\it et al.}
discovered that the spin-flop transition is intimately related to
the magnetoresistance effect \cite{Lavrov}. Similar spin-charge
coupling was also found in a separate study on lightly-doped
Nd$_{1.975}$Ce$_{0.025}$CuO$_{4}$ \cite{li}. The critical field
for the transition from a noncollinear spin structure to a
collinear one increases with decreasing temperature. For lightly
electron-doped $\rm Pr_{1.29}La_{0.7}Ce_{0.01}CuO_{4\pm\delta}$,
the critical field increases from 0.5-T at 150 K to 2-T at 5 K
\cite{Lavrov}.

\begin{figure}
\includegraphics[keepaspectratio=true,width=0.8\columnwidth,clip]{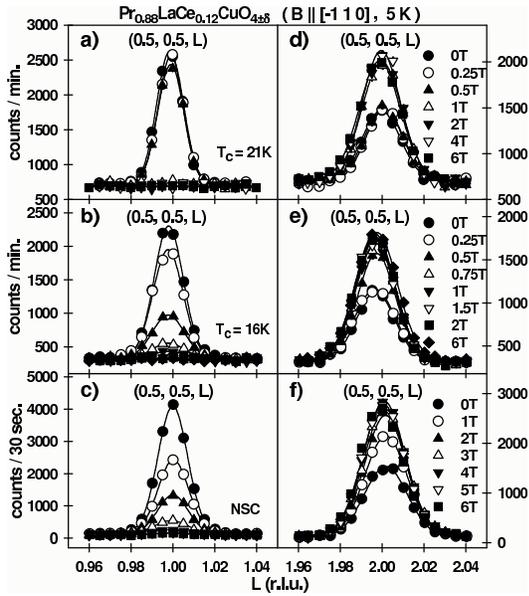}
\caption{Magnetic field dependence of the scattering around
$(0.5,0.5,L)$ at 5 K, a temperature well below the perspective
$T_N$'s for various SC and NSC PLCCO samples. a-c) $L$-scans
around $(0.5, 0.5, 1)$ d-f) $L$-scans around $(0.5, 0.5, 2)$ for
$T_c=21$ K, 16 K and NSC PLCCO samples. The $T_c=21$ K sample
shows an abrupt spin-flop transition and the $T_c=16$ K and the
NSC samples display a more gradual spin-flop transition. The solid
lines are Gaussian fits to the data. }
\end{figure}

We systematically investigated the effect of an in-plane magnetic
field on the spin structure of PLCCO because superconductivity is
not much affected in this geometry. The magnetic structure factors
at $(0.5,0.5,L)$ for the noncollinear and collinear phases of
PLCCO are summarized in Table I. In the collinear spin structure,
the magnetic peaks are disallowed at $(0.5, 0.5, L)$ with $L =$
odd integer. We probed the $(0.5, 0.5, 1)$ and $(0.5, 0.5,2)$
magnetic peaks as a function of the field at low temperature.
Figure 8 summarizes the effect of a ${\bf
B}$$\,\parallel\,$$[\bar{1},1,0]$ field on the $(0.5, 0.5, 1)$ and
$(0.5, 0.5, 2)$ magnetic peaks in different samples of PLCCO at 5
K. The solid lines are Gaussians fits to the data. The $c$-axis
coherence lengths are resolution limited and about 660 \AA\ for
all three samples estimated using
$CL=[\sqrt{2\ln(2)}/\pi](c/\sigma)$. Therefore, the residual AF
order at $(0.5,0.5,L)$ ($L=1$, 2, $\cdots$) in underdoped and NSC
PLCCO are 3D and different from the quasi-2D diffusive scattering
at $(0.5,0.5,0)$.

For the $T_c = 21$ K PLCCO, a 1-T applied field diminishes the
$(0.5,0.5,1)$ and enhances the $(0.5,0.5,2)$ magnetic reflections,
thus indicating that the magnetic structure has been transformed
from the noncollinear to collinear structure by the field (Figs.
7a-b, 8a and 8d). This sharp transition in the $T_c = 21$ K PLCCO
is surprising because spin-flop transitions for ${\bf
B}$$\,\parallel\,$$[\bar{1},1,0]$ fields in lightly-doped $\rm
Pr_{1.29}La_{0.7}Ce_{0.01}CuO_{4\pm\delta}$ \cite{Lavrov},
Nd$_{1.975}$Ce$_{0.025}$CuO$_4$ \cite{li}, Nd$_2$CuO$_4$
\cite{Skanthakumar}, and Pr$_2$CuO$_4$ \cite{sumarlin} are more
gradual and continuous.

\begin{figure}
\includegraphics[keepaspectratio=true,width=0.8\columnwidth,clip]{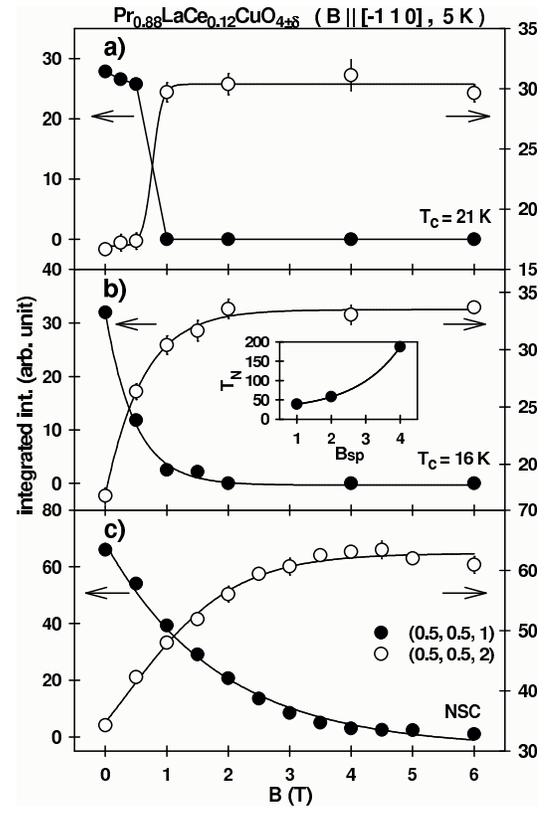}
\caption{Integrated intensities of $(0.5, 0.5, L)$ as a function
of a magnetic field for a) $T_c=21$ K, b) $T_c=16$ K , c) NSC
samples at 5 K. The inset in b) shows that the sample with higher
N\'{e}el temperature requires a higher critical field to induce
the spin-flop transition.}
\end{figure}

To understand the evolution of such behavior as PLCCO is
transformed from a superconductor to an antiferromagnet, we
performed similar measurements on the $T_c = 16$ K and the NSC
PLCCO. The outcome (Figs. 8b, e, c, and f) suggests that spin-flop
transitions become sharper as superconductivity is developed with
increasing $T_c$ and decreasing $T_N$. Figure 9 summarizes the
integrated intensities of $(0.5,0.5,1)$ and $(0.5,0.5,2)$ as a
function of increasing field for the three PLCCO samples at 5 K.
The inset in Fig. 9b shows that a higher N\'{e}el temperature
$T_N$ requires a larger critical field \textbf{B}$_{sp}$ for the
spin-flop transition. \textbf{B}$_{sp}$ therefore decreases with
decreasing $T_N$ and reducing Cu moment.

\begin{figure}
\includegraphics[keepaspectratio=true,width=0.8\columnwidth,clip]{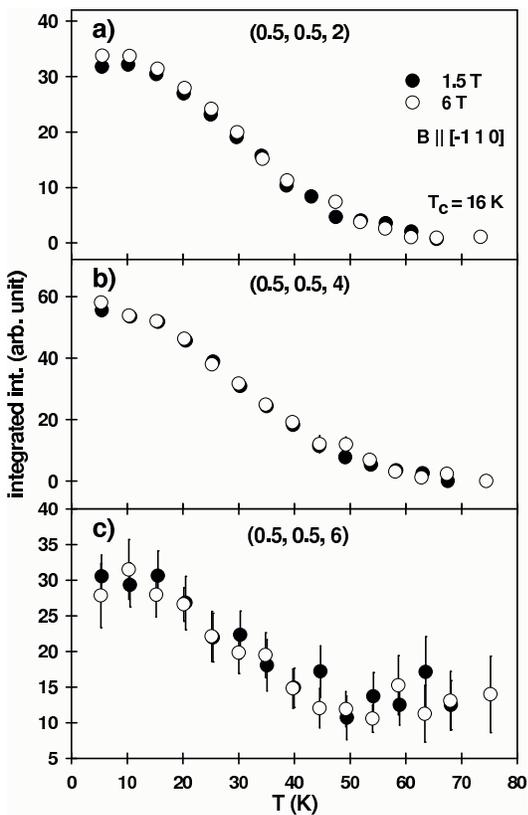}
\caption{The temperature dependence of the integrated magnetic
Bragg intensities for $(0.5,0.5,L)$ with $L=2,4,6$ at 1.5-T and
6-T for the $T_c=16$ K PLCCO. The spin flop transition should have
occurred above 1.5-T (see Fig. 9b). The negligible changes of the
integrated intensities indicate that an in-plane magnetic field of
6-T does not affect the Cu and Pr moment. It only transforms the
magnetic structure from a noncollinear to a collinear one (Fig.
7).}
\end{figure}

From the magnetic structure factor calculations in Table I, we
find that the ratio of integrated intensities of $(0.5, 0.5,
L={\rm even})$ at collinear and noncollinear states is
$[F_{c}(0.5,0.5,L={\rm even})/F_{nc}(0.5,0.5,L={\rm even})]^2=2$
if the Pr and Cu moments under field are the same as that at
zero-field. Inspection of Figs. 8 and 9 for $(0.5,0.5,2)$ shows
that this is indeed the case. Therefore, a moderate magnetic field
($\leq6$-T) that causes spin-flop transition does not induce
additional moments on Cu or Pr sites in PLCCO.

To further determine the effect of a magnetic field on Pr/Cu
moments after the spin-flop transition, we measure the temperature
dependence of the scattering at $(0.5,0.5,L)$ ($L=2$, 4, 6) for
fields just above \textbf{B}$_{sp}$ and at 6-T. This comparison
allows us to determine whether application of additional field
induces Cu/Pr moments in the collinear state of PLCCO after the
spin-flop transition. Inspection of the data in Fig. 10 reveals no
appreciable difference between 1.5-T and 6-T in the temperature
range probed, consistent with the notion that a 6-T in-plane field
does not induce additional Cu/Pr moments for the $T_c = 16$ K
PLCCO.

\begin{figure}
\includegraphics[keepaspectratio=true,width=0.8\columnwidth,clip]{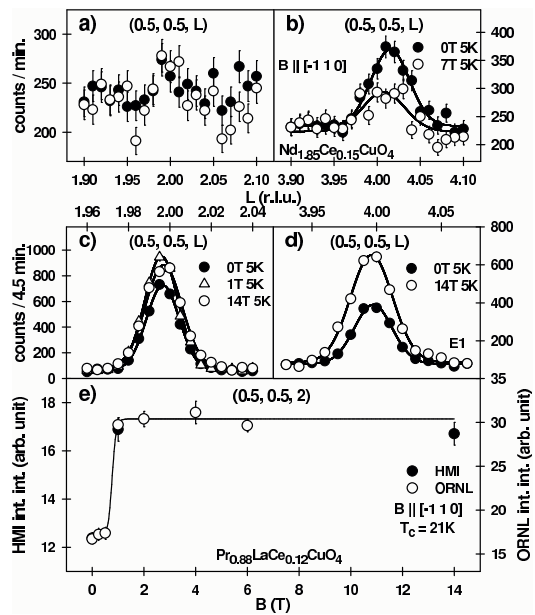}
\caption{The comparison of the influence of an in-plane field on
the magnetic Bragg peaks of PLCCO and NCCO. a-b) low-temperature
$L$ scans across the magnetic peaks at $(0.5, 0.5, 2)$ and $(0.5,
0.5, 4)$ in NCCO (from Ref. \cite{Kang}). c-d) Similar $L$ scans
across $(0.5, 0.5, 2)$ and $(0.5, 0.5, 4)$ in PLCCO. The intensity
differences between NCCO and PLCCO is due the induced moment on
Nd$^{3+}$ in NCCO. e) integrated intensity of (0.5, 0.5, 2) as a
function of a field. Right after spin-flop transition, the
integrated intensity shows no field dependence up to 14-T. This
confirms that Pr moment is not induced by a field up to 14-T. }
\end{figure}

To see if a larger in-plane field will affect the Cu/Pr moment, we
measured the field-dependence of the $(0.5,0.5,2)$ and
$(0.5,0.5,4)$ reflections up to 14-T for the $T_c = 21$ K PLCCO at
HMI. The outcome, shown in Figs. 11c-e, clearly indicates that
while a 1-T in-plane field enhances the intensity of the
$(0.5,0.5,2)$ peak from that at zero-field as expected from the
spin-flop transition, further increasing the applied field to 14-T
does not induce additional changes in the integrated intensity of
the $(0.5,0.5,2)$ peak (Fig. 11c). Therefore, a 14-T field does
not alter the Cu/Pr moments within the uncertainty of our
measurements. This is quite different from the effect of an
in-plane field on Nd$_{1.85}$Ce$_{0.15}$CuO$_4$, where the $(0.5,
0.5, 2)$ reflection shows no intensity differences and the $(0.5,
0.5, 4)$ intensity decreases under a 7-T field (Figs. 11a-b)
\cite{Kang}. For Nd$_{1.85}$Ce$_{0.15}$CuO$_4$, a 7-T ${\bf
B}$$\,\parallel\,$$[\bar{1} ,1,0]$ magnetic field not only causes
a spin-flop transition but also induces magnetic moments on the Nd
sites. The significant intensity drop at $(0.5, 0.5, 4)$ under 7-T
arises because of the larger Nd moment contribution to the
magnetic structure factor (Table I, for NCCO the Nd moment has
opposite direction to Pr moment). Since PLCCO has negligible
field-induced moment contribution from Pr, field-induced effects
here should reveal the inherent Cu$^{2+}$ spin correlations.

\subsection{Magnetic field effect on the nonsuperconducting (NSC) Pr$_{0.88}$LaCe$_{0.12}$CuO$_{4\pm\delta}$
and the cubic (Pr,La,Ce)$_2$O$_3$ impurity phase}

Since the discovery of a $c$-axis magnetic field-induced effect in
Nd$_{1.85}$Ce$_{0.15}$CuO$_4$ \cite{Kang}, there has been much
debate concerning the origin of the field-induced effect. While we
argue that the observed effects are partially intrinsic to
Nd$_{1.85}$Ce$_{0.15}$CuO$_4$ \cite{Matsuura,Kang1,Matsuura1},
Mang {\it et al.} \cite{Mang,Mang1} suggest that all the observed
field-induced effect in Ref. \cite{Kang} can be explained by
paramagnetic scattering from the impurity (Nd,Ce)$_2$O$_3$ phase.
Since there will always be paramagnetic scattering from
(Nd,Ce)$_2$O$_3$ under the influence of a magnetic field in SC
NCCO, it is more productive to study $n$-type superconductors
where the impurity phase has no field-induced effect. SC PLCCO is
a good candidate because Pr$^{3+}$ in the impurity phase
(Pr,La,Ce)$_2$O$_3$ has a nonmagnetic singlet ground state. To
reveal the intrinsic magnetic field effect on Cu$^{2+}$ magnetism
in SC PLCCO, one must understand the field-induced effect on AF
order in NSC PLCCO and impurity (Pr,La,Ce)$_2$O$_3$.

\begin{figure}
\includegraphics[keepaspectratio=true,width=0.8\columnwidth,clip]{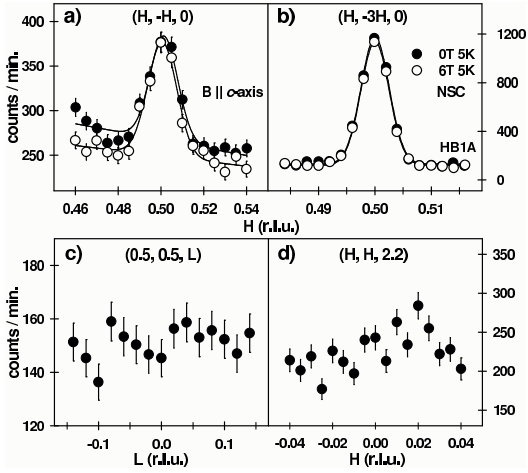}
\caption{a-b) Magnetic field dependent scattering around
$(0.5,-0.5,0)$ and $(0.5,-1.5,0)$ of NSC PLCCO in the $[H,K,0]$
scattering plane. Radial scans across a) $(0.5, -0.5, 0)$ and b)
$(0.5, -1.5, 0)$ at 5 K with zero and 6-T fields, respectively.
The magnetic field is applied along the $c$-axis. While the
scattering at $(0.5,-1.5,0)$ is mostly magnetic from the
noncollinear AF order (Fig. 2a), the peak at $(0.5,-0.5,0)$ does
not arise from (Pr,La,Ce)$_2$O$_3$ but is diffusive structural
scattering of PLCCO similar to those seen in the SC
Nd$_{1.85}$Ce$_{0.15}$CuO$_4$ \cite{Matsuura}. This conclusion is
confirmed by c) $[0.5,0.5,L]$ and d) $[H,H,2.2]$ scans across the
impurity positions of NSC PLCCO in the $[H,H,L]$ scattering plane.
}
\end{figure}

To determine the effect of a $c$-axis aligned magnetic field on
the AF order of NSC PLCCO, we use the reoxygenated sample
($T_N=186$ K) aligned in the $[H,K,0]$ scattering plane. Figures
12a and b show radial scans across $(0.5,-0.5,0)$ and
$(0.5,-1.5,0)$ at zero and 6-T $c$-axis aligned fields. To within
the error of the measurements, a magnetic field of 6-T has no
effect on either peak at 5 K although there are small differences
in the background scattering of Fig. 12a under field with unknown
origin. Since a 6-T field along the $c$-axis has no effect on NSC
Nd$_{1.85}$Ce$_{0.15}$CuO$_4$ \cite{Matsuura}, we conclude that
such field does not affect the AF structure or induce moments in
NSC PLCCO or Nd$_{1.85}$Ce$_{0.15}$CuO$_4$.

Next, we consider the influence of a magnetic field on the
impurity (Pr,La,Ce)$_2$O$_3$ phase. To separate the impurity phase
from PLCCO, we aligned the SC PLCCO single crystals in the
$[H,H,L]$ zone and applied a magnetic field along the
$[\bar{1},1,0]$ direction \cite{Matsuura,Mang1}. Since
(Pr,La,Ce)$_2$O$_3$ has essentially a cubic crystal structure with
lattice parameter $\sim$10\% smaller than the $c$-axis lattice
parameter of PLCCO, one can probe the paramagnetic scattering from
the impurity phase by simply looking around the $(0,0,2.2)$
position of PLCCO in the $[H,H,L]$ geometry \cite{Matsuura,Mang1}.
Here, cubic reflections $(2, 0, 0)\rm_c$ and $(0, 0, 2)\rm_c$ from
(Pr,La,Ce)$_2$O$_3$ can be indexed as $(0.5, 0.5, 0)$ and $(0, 0,
2.2)$ in PLCCO Miller indices, respectively. A ${\bf
B}||[\bar{1},1,0]$ field in the CuO$_2$ plane is along the
$[1,0,0]$ direction of the cubic (Pr,La,Ce)$_2$O$_3$, equivalent
to a $c$-axis aligned field in PLCCO which is along the $[0,0,1]$
direction of (Pr,La,Ce)$_2$O$_3$.

\begin{figure}
\includegraphics[keepaspectratio=true,width=0.8\columnwidth,clip]{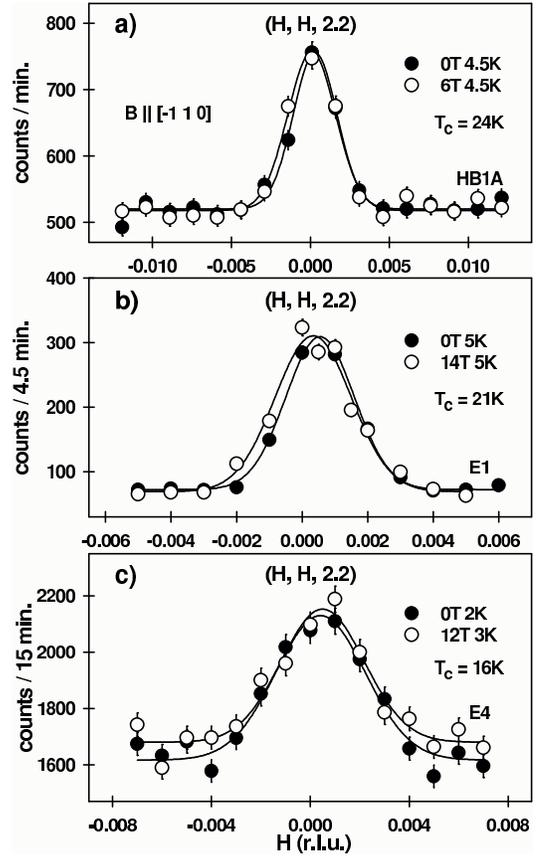}
\caption{Magnetic field effect on the impurity peak position $(0,
0, 2.2)$ at low temperatures. a-c) scans along the $[H,H,2.2]$
direction for the $T_c=24$ K, 21 K, and 16 K PLCCO, respectively.
The magnetic fields are applied along the $[\bar{1},1,0]$
direction, similar to previous work on SC NCCO
\cite{Matsuura,Mang}.}
\end{figure}

The field effect on the (Pr,La,Ce)$_2$O$_3$ impurity phase was
measured in optimally doped and underdoped SC PLCCO samples with
$T_c = 24$ K, 21 K, and 16 K. Following previous work on NCCO
\cite{Matsuura,Mang,Mang1}, we probe the impurity peak positions
$(0,0,2)_{\rm c}$, $(1,1,0)_{\rm c}$, and $(0,0,4)_{\rm c}$ which
correspond to PLCCO Miller indices $(0,0,2.2)$/$(0.5,0.5,0)$,
$(0.5,0,0)$, and $(0,0,4.4)$, respectively. Figure 13 summarizes
the outcome around $(0,0,2.2)$ for the three SC PLCCO samples
investigated. The $[H,H,2.2]$ scans across the impurity position
at $(0, 0, 2.2)$ show no observable field-induced effect on the
impurity phase up to 14-T (Figs. 13a-c). Additional measurements
of the temperature dependent scattering around $(0,0,4.4)$ (Fig.
14a) and the $c$-axis field-dependent scattering around
$(0.5,0,0)$ (Fig. 14b) on the $T_c=21$ K sample confirm that
Pr$^{3+}$ ions in (Pr,La,Ce)$_2$O$_3$ have a nonmagnetic singlet
ground state and cannot be polarized by a 14.5-T field. In
contrast, Nd$^{3+}$ ions in the (Nd,Ce)$_2$O$_3$ impurity phase
can be easily polarized by an applied field. Therefore, we can
avoid the complication of field-induced paramagnetism from the
impurity phase by studying PLCCO and the outcome should
unambiguously reveal the intrinsic properties of Cu$^{2+}$
magnetism in SC electron-doped materials.

\begin{figure}
\includegraphics[keepaspectratio=true,width=0.8\columnwidth,clip]{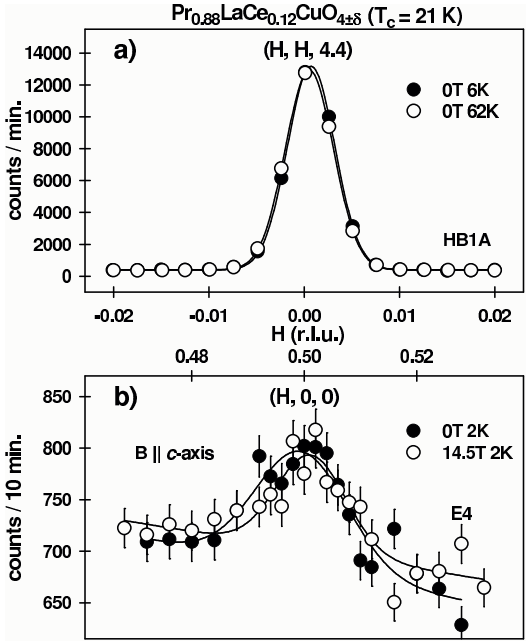}
\caption{a) $[H,H,4.4]$ scan through $(0, 0, 4.4)$ position for
impurity peak $(0, 0, 4)\rm_c$ at low and high temperatures. The
temperature independent scattering indicates that the impurity
peak has no magnetic component. b) Radial $[H,0,0]$ scan across
$(0.5, 0, 0)$ of the impurity peak $(1, 1, 0)\rm_c$ at zero and
14.5-T $c$-axis aligned fields \cite{Matsuura1}. It shows no
field-induced effect, indicating that the impurity phase cannot be
polarized by a 14.5-T field. }
\end{figure}

\subsection{Effect of a $c$-axis aligned magnetic field
on SDW and AF orders in superconducting (SC)
Pr$_{0.88}$LaCe$_{0.12}$CuO$_{4\pm\delta}$}

In this section, we describe the effect of a $c$-axis aligned
magnetic field on SC PLCCO samples. Since the previous section
confirmed that the (Pr,La,Ce)$_2$O$_3$ impurity phase does not
respond to an applied magnetic field, any field-induced effect on
SC samples must be intrinsic to PLCCO. For the experiment, we
aligned crystals in the $[H,K,0]$ scattering plane and applied
magnetic fields along the $c$-axis. Figure 15 shows the radial
$[H,H,0]$ and $[H,3H,0]$ scans around $(0.5,0.5,0)$ and
$(0.5,1.5,0)$ positions, respectively, for the SC $T_c = 24$ K, 21
K, and 16 K samples. In all cases, we carefully applied the
magnetic field above $T_c$ and field-cooled the samples to below
$T_c$. For optimally doped PLCCO ($T_c = 24$ K), a magnetic field
of 6.5-T at 4.5 K does not affect the scattering at $(0.5,0.5,0)$
and $(0.5,1.5,0)$ (Fig. 15a and 15e), consistent with an earlier
report on an overdoped PLCCO ($T_c=16$ K) \cite{fujitaprl}. Note
that in this case the peaks at both positions arise mostly from
superlattice and (Pr,La,Ce)$_2$O$_3$ and there are no static AF
orders in the sample (Fig. 6a and 6b) \cite{daiprl}.

\begin{figure}
\includegraphics[keepaspectratio=true,width=0.8\columnwidth,clip]{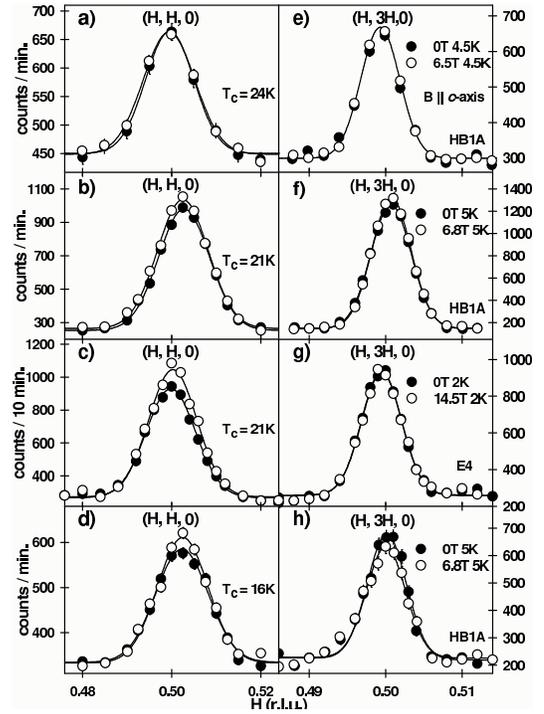}
\caption{The effect of a $c$-axis aligned magnetic field on
various SC PLCCO samples at low temperatures. The data below 7-T
were taken at HB-1A of HFIR while higher field data were collected
on E4 of HMI. The $[H,H,0]$ scans around $(0.5, 0.5, 0)$ for a)
$T_c=24$ K, b,c) 21 K, and d) 16 K PLCCO samples at specified
temperatures. e-h) $[H,3H,0]$ scans around $(0.5, 1.5,0)$ in the
same experimental setup as in a-d).}
\end{figure}

Underdoped SC samples ($T_c = 21$ K and 16 K), however, show clear
low-temperature field-induced intensity gain at the diffusive SDW
position $(0.5,0.5,0)$ (Figs. 15b-d), but not at the 3D AF ordered
position $(0.5,1.5,0)$ (Figs. 15f-h). In particular, the
field-induced intensity appears to increase with increasing
magnetic field for the $T_c=21$ K PLCCO (Figs. 15b and c). On
warming to temperatures above $T_c$ but below $T_N$, the
field-induced effect disappears (Fig. 16). Figure 17 shows the
temperature dependence of the scattering at $(0.5,0.5,0)$ and
$(0.5,1.5,0)$ for the $T_c = 21$ K and 16 K samples at different
fields. For the AF 3D ordered Bragg peak $(0.5,1.5,0)$, the
intensity simply drops with increasing temperature and shows
essentially no difference between field-on and field-off. On the
other hand, scattering at $(0.5,0.5,0)$ shows clear
low-temperature enhancement under field that vanishes at high
temperatures. Although most of the field-induced effects occur at
temperatures below the zero-field $T_c$ as marked, we were unable
to map out the detailed temperature-field dependence of the effect
due to weakness of the signal. This is similar to the
field-induced effect on incommensurate SDW in LSCO \cite{Lake1}
and La$_2$CuO$_{4+y}$ \cite{Khaykovich1}.

\begin{figure}
\includegraphics[keepaspectratio=true,width=0.8\columnwidth,clip]{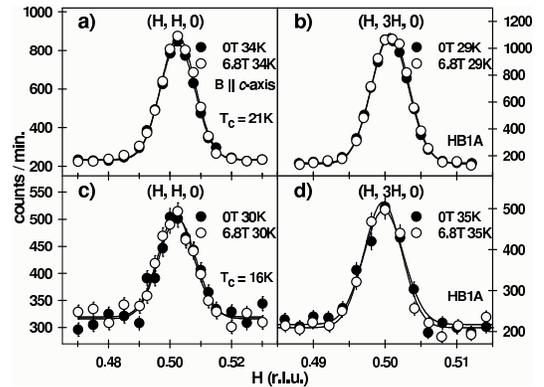}
\caption{The field dependence of the scattering across $(0.5, 0.5,
0)$ and $(0.5, 1.5, 0)$ at temperatures above $T_c$ and below
$T_N$. a-b) for the $T_c=21$ K sample. c-d) for the $T_c=16$ K
sample. No field-induced effects are observed.}
\end{figure}

\begin{figure}
\includegraphics[keepaspectratio=true,width=0.8\columnwidth,clip]{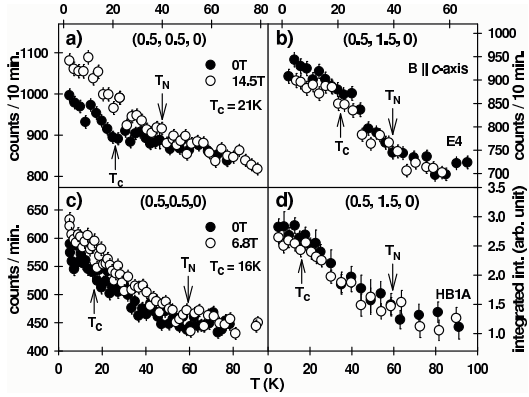}
\caption{Temperature dependence of the scattering at the $(0.5,
0.5, 0)$ and $(0.5, 1.5, 0)$ positions for a-b) $T_c=21$ K and
c-d) $T_c=16$ K samples at zero and finite $c$-axis aligned
magnetic fields. Clear low-temperature field-induced effects are
seen at $(0.5,0.5,0)$ but not at $(0.5,1.5,0)$ for both the
$T_c=21$ K and 16 K PLCCO samples. }
\end{figure}

Finally, we carried out a series of $[H,H,0]$ and $[H,3H,0]$ scans
across $(0.5,0.5,0)$ and $(0.5,1.5,0)$ respectively for the
$T_c=21$ K sample to determine the low-temperature
field-dependence of the scattering (Fig. 18). The open circles
show the ORNL data taken at 5 K for fields up to 6.8-T while the
filled circles are HMI data at 2 K for fields up to 14.5-T. The
integrated intensity of the SDW at each field is computed by
fitting the raw data with a Gaussian on a linear background (See
Fig. 15). By combining ORNL and HMI data, we find that the
field-induced effect at $(0.5,0.5,0)$ increases linearly with
increasing magnetic field up to 13.5 T and may saturate at higher
fields (Fig. 18a). On the other hand, the 3D AF residual order
$(0.5,1.5,0)$ peak of the $T_c=21$ PLCCO shows no observable
field-induced effect for fields up to 14.5-T (Fig. 18b).

\subsection{Magnetic field-induced anisotropy on the $T_c=21$ K superconducting (SC) Pr$_{0.88}$LaCe$_{0.12}$CuO$_{4\pm\delta}$}

The discovery of a $c$-axis field-induced effect unrelated to the
(Pr,La,Ce)$_2$O$_3$ impurity phase in underdoped PLCCO is
suggestive but not a proof that such effect is related to the
suppression of superconductivity, as one might argue that the AF
phase in the SC PLCCO is somehow different from those in NSC
samples \cite{Mang1}. One way to check whether the observed effect
is indeed related to the suppression of superconductivity is to
determine its field directional dependence. Since high-$T_c$
superconductors are layered materials, a magnetic field aligned
along the $c$-axis suppresses superconductivity much more
dramatically than the same field parallel to the CuO$_2$ planes.

\begin{figure}
\includegraphics[keepaspectratio=true,width=0.8\columnwidth,clip]{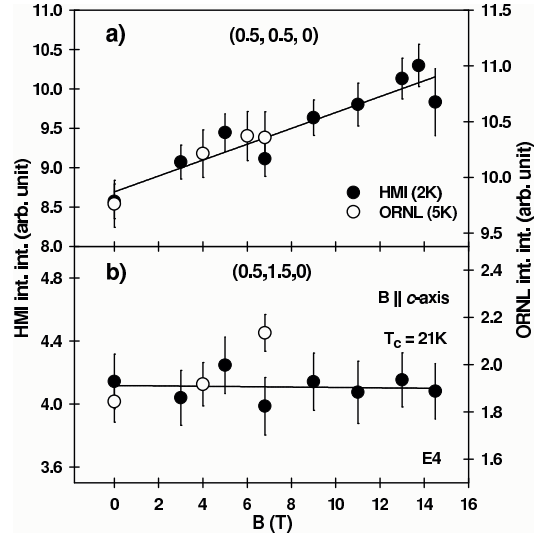}
\caption{a) Integrated intensity of $(0.5, 0.5, 0)$ as a function
of increasing field. Open circles are data from ORNL at 5 K and
filled circles are HMI data at 2 K. b) integrated intensity of
$(0.5, 1.5, 0)$ is independent of applied field up to 14.5-T. }
\end{figure}

To compare the in-plane and $c$-axis magnetic field-induced
effect, we focus on the SDW at $(0.5,0.5,0)$ for the $T_c=21$ K
PLCCO. Figure 19 shows scans along the $[H,H,0]$ and $[0.5,0.5,L]$
directions across $(0.5,0.5,0)$ at zero and 6-T {\bf
B}$||[\bar{1},1,0]$ in-plane field. There is no observable
field-induced effect, in contrast to $c$-axis field data across
$(0.5,0.5,0)$ (Fig. 15b-c). On increasing the in-plane applied
field to 14-T, we again failed to observe any field-induced effect
around $(0.5,0.5,0)$ at 5 K (Figs. 20b and c). For comparison, we
replot the data of Fig. 15c in Fig. 20a. It is clear that a
$c$-axis aligned field enhances scattering at $(0.5,0.5,0)$ while
the same field parallel to the CuO$_2$ planes does not. Since a
14-T in-plane field only induces the spin-flop transition and does
not affect the Cu/Pr moments of the residual AF phase, our
observation of an anisotropic field effect in the $T_c=21$ K PLCCO
is the most direct evidence that the enhancement of SDW order is
related to the suppression of superconductivity.

\begin{figure}
\includegraphics[keepaspectratio=true,width=0.8\columnwidth,clip]{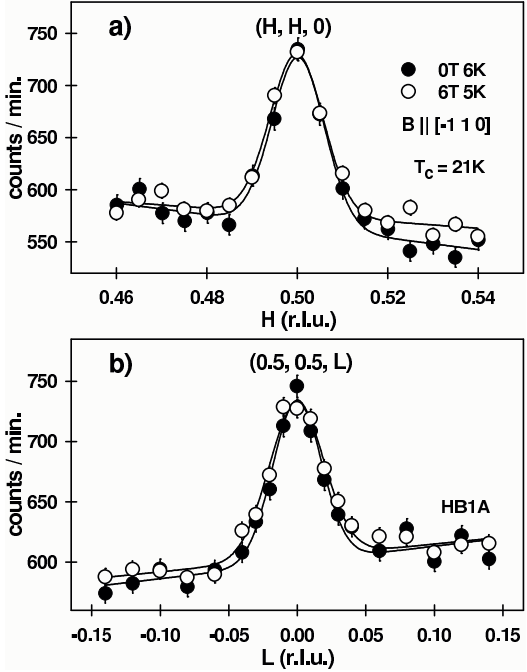}
\caption{The influence of an in-plane magnetic field on the SDW
order in the $T_c=21$ K SC PLCCO. a) $[H,H,0]$ scan around $(0.5,
0.5, 0)$ for zero and 6-T at 5 K. b) $[0.5,0.5,L]$ scan around
$(0.5, 0.5, 0)$. A 6-T in-plane magnetic field has no observable
effect on the SDW order at 5 K. The $c$-axis coherence length of
$(0.5, 0.5, 0)$ is about 200 \AA. }
\end{figure}

The presence of a field-induced effect at the $(0.5,0.5,0)$ SDW
position but not at the AF $(0.5,1.5,0)$ Bragg position suggests
that Cu spins contributing to the diffusive SDW cannot arise from
the same Cu spins giving the 3D AF moments. While similar
field-induced enhancement was also observed in the case of
optimally doped NCCO \cite{Kang}, there are important differences
between PLCCO and NCCO. First, optimally doped PLCCO has no
residual AF order while 3D AF order coexists with
superconductivity in NCCO even for samples with the highest $T_c$
\cite{note2}. In addition, there are no detailed studies on how
NCCO is transformed from an AF insulator to an optimally doped
superconductor. As a consequence, it is unclear how to compare
PLCCO directly with NCCO. Second, a $c$-axis aligned field applied
on the optimally doped NCCO not only enhances the magnetic signal
at $(0.5,0.5,0)$, but also at AF Bragg positions such as
$(0.5,1.5,0)$ and $(0.5,0.5,3)$ \cite{Matsuura}. For underdoped
PLCCO, a 14-T field has no observable effect on $(0.5,1.5,0)$ AF
3D order. Finally, a $c$-axis aligned field enhances the intensity
of the Bragg peak at $(1,1,0)$ for both NSC and SC NCCO
\cite{Matsuura} while a similar field has no observable effect on
any nuclear Bragg peaks in NSC and SC PLCCO. This difference must
arise from the polarization of the Nd$^{3+}$ ions in NCCO and
therefore may not be intrinsic to the physics of electron-doped
copper oxides.

\section{Discussion}

We use neutron scattering to study the phase transition of PLCCO
from a long-range ordered antiferromagnet to a high-$T_c$
superconductor without static AF order \cite{daiprl}. In the
underdoped regime, we observe the coexistence of a quasi-2D SDW
and a 3D AF order in the superconducting state. The 3D AF order
has the same noncollinear magnetic structure as that in undoped
Pr$_2$CuO$_4$ (Fig. 2a). Since the noncollinear spin structure
does not allow magnetic Bragg scattering at $(0.5, 0.5, 0)$, it is
interesting to ask whether the Cu spins giving rise to the 3D AF
order also contribute to the diffusive quasi-2D scattering.

In one picture, the annealing process necessary for producing
superconductivity in PLCCO may also induce macroscopic oxygen
inhomogeneities, giving rise to mesoscopic separation between the
SC and AF NSC phases. In the AF NSC phase, a weak random rotation
of the magnetic moment from one plane to another induces a
long-range spin order in the plane, but along the $c$-axis one
observes simultaneously a Bragg peak and a diffusive scattering.
If both the diffusive $(0.5,0.5,0)$ and 3D AF $(0.5,1.5,0)$
signals are from the same Cu$^{2+}$ spins macroscopically phase
separated from nonmagnetic SC regions, one would expect that
application of a $c$-axis aligned magnetic field will have no (or
the same) effect on both the 2D and 3D scattering. Instead, we
find a clear $c$-axis field-induced effect at $(0.5,0.5,0)$ but
not at $(0.5,1.5,0)$ (see section III.D). Therefore, Cu$^{2+}$
spins contributing to the diffusive scattering cannot arise from
the same Cu$^{2+}$ spins giving 3D AF moments.

\begin{figure}
\includegraphics[keepaspectratio=true,width=0.8\columnwidth,clip]{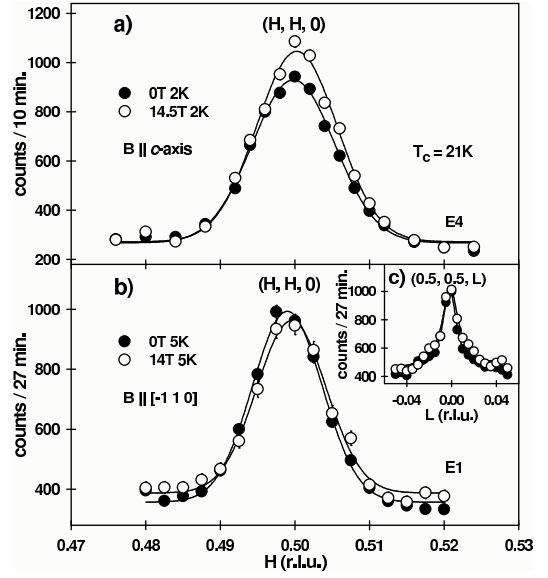}
\caption{Anisotropic field-induced effect on the SDW order of the
$T_c=21$ K PLCCO by a $c$-axis aligned magnetic field and an
$ab$-plane magnetic field. a) $[H,H,0]$ scan across $(0.5, 0.5,
0)$ with zero and 14.5-T field at 2 K in the $[H,K,0]$ zone. b)
$[H,H,0]$ scan across $(0.5, 0.5, 0)$ with a ${\bf B}||ab$-plane
magnetic field at zero and 14-T. c) $[0.5,0.5,L]$ scan across
$(0.5, 0.5, 0)$ with zero and 14-T at 5 K. }
\end{figure}

Alternatively, the annealing process may produce local oxygen
distribution fluctuations acting as pinning centers for
microscopic electronic phase separation. In this case, the
quasi-2D AF scattering may arise from weakly correlated CuO$_2$
layers with a checkerboard AF order in the matrix of the 3D AF
state. The observed diffusive commensurate SDW in PLCCO is then
analogous to the incommensurate 2D SDW in hole-doped LSCO and
La$_2$CuO$_{4+y}$ \cite{Kimura,Lee}. When a $c$-axis aligned
magnetic field is applied, the residual 3D AF order is not
disturbed but the diffusive SDW at $(0.5,0.5,0)$ is enhanced at
the expense of superconductivity. To estimate an effective moment
of the 2D SDW order, we assume two models of spin structures in
Fig. 21 that can give magnetic scattering at $(0.5,0.5,0)$. Of
course, we don't know the stability of the proposed spin
structures in the tetragonal PLCCO unit cell and cannot determine
the location of the Cu spins that contribute the quasi-2D and 3D
AF responses because neutrons are a bulk probe. Our purpose of
introducing these models is to estimate an average Cu moment by
normalizing the magnetic intensity at $(0.5,0.5,0)$ to the weak
nuclear $(1,1,0)$ reflection. To estimate the effective Cu$^{2+}$
moment at zero and 14.5-T, we first determine the magnitude of
magnetic scattering at zero field by subtacting the integrated
intensity of $(0.5,0.5,0)$ at high temperatures from that at 2 K.
The total magnetic scattering at 14.5-T is also determined by
subtracting the high temperature nonmagnetic structural peak from
that at 2 K. For $T_c=21$ K PLCCO at 2 K, the integrated magnetic
intensities at zero and 14.5-T are 2.0528 and 3.3411,
respectively. This means that a 14.5-T field enhances the magnetic
scattering at $(0.5,0.5,0)$ by more than 60\%
[$I$(14.5-T)$/I$(0-T)$=1.66$], much larger than the raw data in
Figs. 15 and 16 would suggest. On the same incident beam monitor
count, the integrated intensity of $(1,1,0)$ nuclear Bragg peak is
4547.

\begin{table}
\caption{Magnetic neutron scattering intensity calculations for
proposed spin structures in Fig. 21. The angle $\theta$ is the
scattering angle for the Bragg peak.
$I_{cal}(0.5,0.5,0)/I_{cal}(1,1,0)=I_{ob}(0.5,0.5,0)/I_{ob}(1,1,0)=$
2.7415$\cdot (0.939M_{cu})^2$/5.6531=2.0528/4547 at 0-T, and
3.3411/4547 at 14.5-T using spin structure I. }
\begin{ruledtabular}
\begin{tabular}{cc}
$(H,K,L)$ & $I_{model\ I}$(domains I \& II) \\
\hline
$(0.5,0.5,0)$ & $\frac{1}{2sin2\theta}$ 32.0 ($\gamma$e$^2$/2$m$c$^2$)$^2$ ($f_{cu}M_{cu}$ - 0.88$f_{pr}M_{pr})^2$ \\
$(0.5,1.5,0)$ & $\frac{1}{2sin2\theta}$ 6.4 ($\gamma$e$^2$/2$m$c$^2$)$^2$ ($f_{cu}M_{cu}$ - 0.88$f_{pr}M_{pr})^2$ \\

\hline\hline

$(H,K,L)$ & $I_{model\ II}$(domains I \& II) \\
\hline

$(0.5,0.5,0)$ & $\frac{1}{2sin2\theta}$ 64.0 ($\gamma$e$^2$/2$m$c$^2$)$^2$ ($f_{cu}M_{cu}$ - 0.88$f_{pr}M_{pr})^2$ \\
$(0.5,1.5,0)$ & $\frac{1}{2sin2\theta}$ 12.8 ($\gamma$e$^2$/2$m$c$^2$)$^2$ ($f_{cu}M_{cu}$ - 0.88$f_{pr}M_{pr})^2$ \\

\hline\hline

$(H,K,L)$ & $I_{structural}$ in magnetic unit cell \\
\hline

$(1,1,0)$ & $\frac{1}{sin2\theta}$ 64.0 (0.88$b_{Pr}+b_{La}+0.12b_{Ce}+b_{Cu}-4b_{O})^2$ \\

\end{tabular}
\end{ruledtabular}
\end{table}

Assuming each collinear model has two equally populated domains as
depicted in Fig. 21 and negligible Pr moment contribution, we can
calculate the expected magnetic intensities of these two models in
Table II. For the collinear model I with domains I/II (Fig.
21a-b), we obtain the Cu moment 0.032 $\mu$$_B$ at 0-T and 0.041
$\mu$$_B$ at 14.5-T. The enhancement of spin correlations appears
mostly at $(0.5, 0.5, 0)$ and the intensity at $(0.5,1.5,0)$ is
about 16 times smaller [$I(0.5, 1.5,0)/I(0.5, 0.5, 0)=0.06$, here
we used $\theta$ calculated for $E_i = E_f = 13.7$ meV.]. This is
consistent with our observable of a weak (or no) field-induced
effect at $(0.5,1.5,0)$. The collinear model II with domains I/II
are essentially the same as that of the model I, except the spin
directions are along the $[1,1,0]$ and $[\bar{1},1,0]$ directions
(Figs. 21c-d). This structure gives Cu moments of 0.022 $\mu_B$ at
0-T and 0.028 $\mu_B$ at 14.5-T. Therefore, the field-induced
enhancement of the Cu moment is smaller ($\sim 0.006$ $\mu_B$)
than that of hole-doped LSCO and LCO in these two models
\cite{Lake1,Khaykovich}.

\begin{figure}
\includegraphics[keepaspectratio=true,width=0.8\columnwidth,clip]{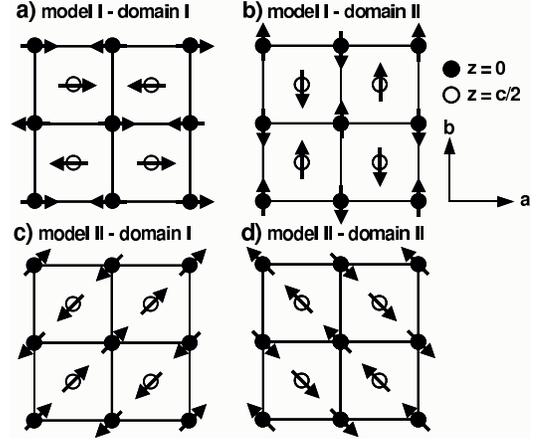}
\caption{Possible Cu$^{2+}$ magnetic structures responsible for
the SDW order. a-b) Collinear spin model I, domains I and II. c-d)
collinear spin model II, domains I and II. In collinear spin
structures, the moments are in the $ab$-plane. The closed and open
circles represent the $z=0$ and $z= c/2$ CuO$_2$ planes,
respectively. The Pr$^{3+}$ spins are not shown, since the induced
moment of Pr$^{3+}$ is negligible in the SC PLCCO samples. Here we
assumed weak (2 or 3 lattices) spin correlations along the
$c$-axis. Experimentally, we probably cannot distinguish such
scattering from that of random 2D sheets in the 3D AF matrix.}
\end{figure}

When as-grown PLCCO is transformed from a long-range ordered
antiferromagnet to a high-$T_c$ superconductor by annealing in
pure Ar, the 3D AF order is degraded and a quasi-2D SDW order
develops with the appearance of superconductivity. As optimal
superconductivity is reached with $T_c=24$ K, both static 3D AF
and diffusive SDW orders vanish. We confirm that the annealing
process necessary for producing superconductivity also induces the
bixbyite (Pr,La,Ce)$_2$O$_3$ as an impurity phase. However, in
contrast to (Nd,Ce)$_2$O$_3$ in SC NCCO, a 14-T field cannot
polarize Pr$^{3+}$ in (Pr,La,Ce)$_2$O$_3$ and this property
therefore allows an unambiguous identification of the intrinsic
field-induced effect on PLCCO. At optimal doping, a 6.5-T $c$-axis
aligned field does not induce static AF order due to the presence
of a spin gap \cite{fujitaprl}. For underdoped materials, a 14-T
$c$-axis field enhances the SDW at $(0.5,0.5,0)$ but has no
observable effect on 3D residual AF order at $(0.5,1.5,0)$. A
similar field applied along the $[\bar{1},1,0]$ direction in the
CuO$_2$ plane only induces a spin-flop transition but has no
effect on the SDW order. This is the most direct evidence that the
enhancement of SDW is associated with the suppression of
superconductivity.

\begin{table}
\caption{Magnetic structure factor calculation for noncollinear
spin structure of NCCO (type I/III) and PLCCO. }
\begin{ruledtabular}
\begin{tabular}{cc}
$(H,K,L)$ & $(F)^2$ for NCCO type I/III \\
\hline
$(0.5,1.5,0)$ & 25.6 ($\gamma$e$^2$/2$m$c$^2$)$^2$ ($f_{cu}M_{cu}$ + 1.85$f_{nd}M_{nd})^2$ \\
$(0.5,0.5,3)$ & 32.0 ($\gamma$e$^2$/2$m$c$^2$)$^2$ ($f_{cu}M_{cu}$ + 1.7426$f_{nd}M_{nd})^2$ \\

\hline\hline

$(H,K,L)$ & $(F)^2$ for PLCCO \\
\hline

$(0.5,1.5,0)$ & 25.6 ($\gamma$e$^2$/2$m$c$^2$)$^2$ ($f_{cu}M_{cu}$ $-$ 0.88$f_{pr}M_{pr})^2$ \\
$(0.5,0.5,3)$ & 32.0 ($\gamma$e$^2$/2$m$c$^2$)$^2$ ($f_{cu}M_{cu}$ $-$ 0.8196$f_{pr}M_{pr})^2$ \\

\end{tabular}
\end{ruledtabular}
\end{table}

While our results for optimally doped PLCCO are consistent with
that for an overdoped PLCCO by Fujita {\it et al.}
\cite{fujitaprl}, we cannot confirm their results for the $x=0.11$
($T_c=25$ K) PLCCO. In particular, the 3D AF order has no
observable field dependence in our underdoped samples. On the
other hand, it would also be interesting to check if their
$x=0.11$ PLCCO sample has a field-induced effect at $(0.5,0.5,0)$.
At present, it is unclear how to reconcile our results with that
of Fujita {\it et al.} \cite{fujitaprl}. Perhaps the differences
in Ce concentration and/or post annealing oxygen treatment plays
an important role in determining the properties of these
materials. In any case, what is clear is that underdoped PLCCO has
a field-induced effect unrelated to the impurity phase, and such
an effect appears to be associated with the suppression of
superconductivity \cite{Chen}.

To determine the influence of an applied field on the residual 3D
AF order in SC electron-doped cuprates, one must understand the
differences in the field-induced effect between optimally doped
NCCO and underdoped PLCCO. The optimally doped NCCO shows clear
$c$-axis field-induced effects at the 3D AF ordering positions
such as $(0.5, 1.5,0)$ and $(0.5,0.5,3)$. On the other hand, we
were unable to find clear field-induced effects at $(0.5, 1.5,0)$
for underdoped PLCCO. This difference may arise from the different
rare-earth moment contributions to the magnetic scattering. For
NCCO, the structure factor calculations in Table III show that the
enhanced Nd$^{3+}$ moment from the exchange coupling to the
field-induced Cu$^{2+}$ moment contributes constructively to the
$(0.5, 1.5,0)$ and $(0.5,0.5,3)$ reflections. On the other hand,
the Pr$^{3+}$ induced moment contributes destructively to the
Cu$^{2+}$ moment in PLCCO. As a consequence, any field-induced
Pr$^{3+}$ moment should reduce the scattering intensity at $(0.5,
1.5,0)$ for PLCCO while a small Cu$^{2+}$ moment enhancement will
induce large intensity enhancement at $(0.5, 1.5,0)$ and
$(0.5,0.5,3)$ for NCCO. Since SC PLCCO has a negligible
field-induced Pr$^{3+}$ moment, a small field-induced enhancement
in the Cu$^{2+}$ moment may not be easily observable at $(0.5,
1.5,0)$.

Our results can be thought of as analogous to that in hole-doped
LSCO. When superconductivity first emerges in LSCO with increasing
Sr-doping, quasi-2D incommensurate SDW peaks also appear
spontaneously to coexist with superconductivity as shown in Fig.
1b \cite{wakimoto,fujitalsco,matsudalsco}. As LSCO becomes an
optimally doped superconductor, the static SDW is replaced by a
spin gap and incommensurate spin fluctuations at energies above
the gap \cite{kyamada}. A $c$-axis aligned magnetic field that
strongly suppresses superconductivity also enhances the static SDW
order in underdeopd LSCO \cite{Katano,Lake1} and induces states
within the gap in optimally doped LSCO \cite{lake2,tranquada}.
Very recently, Khaykovich {\it et al.} \cite{khaykovich3} report
the observation of a magnetic-field-induced transition between
magnetically disordered and ordered phases in slightly underdoped
LSCO with $x=0.144$. Here, static incommensurate SDW order can be
induced directly by a field from the sample without zero-field SDW
order.

For electron-doped PLCCO, we also find the simultaneous appearance
of a quasi-2D SDW order and superconductivity, except in this case
the SDW modulations are commensurate with the underlying lattice
(Fig. 1a) \cite{daiprl}. Similar to LSCO, optimally doped PLCCO
also does not exhibit static SDW order. Work is currently underway
to determine the size of the spin gap and spin fluctuations at
high energies. When a $c$-axis aligned magnetic field is applied,
the static commensurate SDW order is enhanced for underdoped PLCCO
but not for optimally doped samples. These results are very
similar to hole-doped LSCO, thus suggesting the universality of
the magnetic properties in both hole- and electron-doped cuprates.
In the coming years, neutron scattering experiments will be
carried out to search for the spin excitations in optimally and
underdoped PLCCO. Our ultimate goal is to determine whether
magnetism plays a fundamental role in controlling the properties
of doped cuprates and the mechanism of high-$T_c$
superconductivity

\section{Summary and conclusions}

We have systematically investigated the effect of a magnetic field
on electron-doped PLCCO materials from an AF insulator to an
optimally doped superconductor. By controlling the annealing
temperatures, we obtain PLCCO samples with different $T_c$'s and
$T_N$'s. When superconductivity first appears in PLCCO, a quasi-2D
SDW order is also induced at $(0.5,0.5,0)$, and both coexist with
the residual 3D AF state. While the annealing process also induces
cubic (Pr,La,Ce)$_2$O$_3$ as an impurity phase similar to
(Nd,Ce)$_2$O$_3$ in SC NCCO, we show that a magnetic field up to
14-T does not induce magnetic scattering in this impurity phase.
To determine whether AF order is a competing ground state for
superconductivity in electron-doped superconductors, we first must
understand the influence of a magnetic field on the residual AF
order for SC and NSC PLCCO without the complication of
superconductivity. We confirm that a $c$-axis aligned magnetic
field has no effect on the AF order in the NSC PLCCO. Because
PLCCO is a layered superconductor with highly anisotropic
$B_{c2}$, superconductivity can be dramatically suppressed by a
${\bf B}||c$-axis field but much less affected by the same field
in the CuO$_2$ plane. Utilizing this property, we determine the
influence of an in-plane magnetic field on the residual AF order
in SC PLCCO. We find that a 14-T field along the $[\bar{1},1,0]$
direction only causes a spin-flop transition but does not induce
additional moments on Cu$^{2+}$ and/or Pr$^{3+}$ site in the SC
PLCCO's. This suggests that the residual AF order in SC PLCCO
behaves in a similar way as AF order in NSC materials.

To study the magnetic field effect on SDW modulations in SC PLCCO,
we conducted experiments in two ways. First, we aligned the
crystal in the $[H,K,0]$ geometry and applied magnetic field along
the $c$-axis. For both $T_c=21$ K and 16 K PLCCO samples, the
$c$-axis aligned field enhances the SDW order at the $(0.5, 0.5,
0)$ position but has no observable effect at the 3D AF order
position $(0.5,1.5,0)$. The field-induced enhancement at $(0.5,
0.5, 0)$ increases with increasing magnetic field up to 13.5-T for
the $T_c=21$ K sample. The $T_c = 16$ K sample also shows a
field-induced effect at $(0.5,0.5,0)$ for measured fields up to
6.8-T. Second, the crystal was aligned in the $[H,H,L]$ scattering
plane and a magnetic field was applied along the $[\bar{1},1,0]$
direction. In this scattering geometry, both $[0.5,0.5,L]$ and
$[H,H,0]$ radial scans around the $(0.5,0.5,0)$ position in the
CuO$_2$-plane show no observable field-induced effect. This
anisotropy of a field effect confirms that the enhancement of SDW
in the $c$-axis field direction is related to the suppression of
superconductivity.

\section{Acknowledgments}

We are grateful to M. Matsuura, Jerel Zarestky, and K. Prokes for
technical assistance with our neutron scattering measurements at
ORNL and HMI, respectively. We would also like to acknowledge
Branton Campbell, M. Matsuura, Stephan Rosenkranz, J. M.
Tranquada, S. C. Zhang, and K. Yamada for helpful discussions.
This work is supported by the U. S. NSF DMR-0139882, DOE No.
DE-AC05-00OR22725 with UT/Battelle, LLC.


\begin{references}
\bibitem{kastner} M. A. Kastner, R. J. Birgeneau, G. Shirane, and Y. Endoh, Rev. Mod. Phys. {\bf 70}, 897 (1998).
\bibitem{sczhang} S. Sachdev and S.-C. Zhang, Science {\bf 295}, 452 (2002).
\bibitem{levi} B. G. Levi, Physics Today {\bf 57}, 24 (2004).
\bibitem{Kimura} H. Kimura, K. Hirota, H. Matsushita, K.Yamada, Y. Endoh, S-H
Lee, C. F. Majkrzak, R. Erwin, M. Greven, Y. S. Lee, M. A.
Kastner, and R. J. Birgeneau, Phys. Rev. B {\bf 59}, 6517 (1999).
\bibitem{Lee} Y. S. Lee, R. J. Birgeneau, M. A. Kastner, Y. Endoh, S.
Wakimoto, K. Yamada, R. W. Erwin, S.-H. Lee, and G. Shirane, Phy.
Rev. B {\bf 60}, 3643 (1999).
\bibitem{kyamada} K. Yamada, C. H. Lee, K. Kurahashi, J. Wada, S. Wakimoto, S. Ueki, H. Kimura,
Y. Endoh, S. Hosoya, G. Shirane, R. J. Birgeneau, M. Greven, M. A.
Kastner, and Y. J. Kim, Phys. Rev. B {\bf 57}, 6165 (1998).
\bibitem{wakimoto} S. Wakimoto, G. Shirane, Y. Endoh, K. Hirota, S. Ueki, K. Yamada, R. J. Birgeneau, M. A. Kastner, Y. S. Lee, P. M. Gehring, and S. H. Lee, Phys. Rev. B {\bf 60}, R769 (1999).
\bibitem{fujitalsco} M. Fujita, K. Yamada, H. Hiraka, P. M. Gehring, S. H. Lee, S. Wakimoto, and G. Shirane,
Phys. Rev. B {\bf 65}, 064505 (2002).
\bibitem{matsudalsco} M. Matsuda, M. Fujita, K. Yamada, R. J. Birgeneau, Y. Endoh, and G. Shirane, Phys. Rev. B
{\bf 65}, 134515 (2002).
\bibitem{Katano} S. Katano, M. Sato, K. Yamada, T. Suzuki, and T. Fukase, Phy.
Rev. B {\bf 62}, R14677 (2000).
\bibitem{Lake1} B. Lake, H. M. Ronnow, N. B. Christensen, G. Aeppli, K.
Lefmann, D. F. McMorrow, P. Vorderwisch, P. Smeibidl, N.
Mangkorntong, T. Sasagawa, M. Nohara, H. Takagi, and T. E. Mason,
Nature {\bf 415}, 299 (2002)
\bibitem{Khaykovich} B. Khaykovich, Y. S. Lee, R. W. Erwin, S.-H. Lee, S. Wakimoto,
K. J. Thomas, M. A. Kastner, and R. J. Birgeneau, Phy. Rev. B {\bf
66}, 014528 (2002).
\bibitem{Khaykovich1} B. Khaykovich, R. J. Birgeneau, F. C. Chou, R. E. Erwin, M. A. Kastner,
S.-H. Lee, Y. S. Lee, P. Smeibidl, P. Vorderwisch, and S.
Wakimoto, Phys. Rev. B {\bf 67}, 054501 (2003).
\bibitem{lake2} B. Lake, G. Aeppli, K. N. Clausen, D. F. McMorrow, K. Lefmann, N. E. Hussey,
N. Mangkorntong, N. Nahara, H. Takagi, and T. E. Mason, Science
{\bf 291}, 1759 (2001).
\bibitem{tranquada} J. M. Tranquada, C. H. Lee, K. Yamada, Y. S. Lee, L. P. Regnault, H. M. Ronnow, Phys. Rev. B {\bf 69}, 174507 (2004).
\bibitem{Lake} B. Lake, K. Lefmann, N. B. Christensen, G. Aeppli, D. F.
McMorrow, P. Vorderwisch, P. Smeibidl, N. Mangkorntong, T.
Sasagawa, M. Nohara, H. Takagi, (unpublished).
\bibitem{Zhang} Shou-Cheng Zhang, Science {\bf 275}, 1089 (1997).
\bibitem{Arovas} Daniel P. Arovas, A. J. Berlinsky, C. Kallin, and Shou-Cheng
Zhang, Phy. Rev. Lett. {\bf 79}, 2871 (1997).
\bibitem{demler} Y. Zhang, E. Demler, and S. Sachdev, Phys. Rev. B {\bf 66}, 094501 (2002).
\bibitem{dhlee} D.-H. Lee, Phys. Rev. Lett. {\bf 88}, 227003 (2002).
\bibitem{ychen} Y. Chen and C. S. Ting, Phys. Rev. B {\bf 65}, 180513 (2002).
\bibitem{skivelson} S. A. Kivelson, D.-H. Lee, E. Fradkin, and O. Oganesyan, Phys. Rev. B {\bf 66}, 144516 (2002).
\bibitem{jzhu} J. Zhu, I. Martin, and A. Bishop, Phys. Rev. Lett. {\bf 89}, 067003 (2002).
\bibitem{dhz} E. Demler, W. Hanke, and S.-C. Zhang, Rev. Mod. Phys. {\bf 76}, 909 (2004).
\bibitem{dainature} Pengcheng Dai, H. A. Mook, G. Aeppli, S. M. Hayden, and F. Do$\rm\breve{g}$an, Nature
(London) {\bf 406}, 965 (2000).
\bibitem{mookprb} H. A. Mook, Pengcheng Dai, S. M. Hayden, A. Hiess, J. W. Lynn, S.-H. Lee,
and F. Do$\rm\breve{g}$an, Phys. Rev. B {\bf 66}, 144513 (2002).
\bibitem{hidaka} Y. Hidaka and M. Suzuki, Nature (London) {\bf 338}, 635 (1989).
\bibitem{fournier} P. Fournier, P. Mohanty, E. Maiser, S. Darzens, T. Venkatesan,
C. J. Lobb, G. Czjzek, R. A. Webb, and R. L. Greene, Phys. Rev.
Lett. {\bf 81}, 4720 (1998).
\bibitem{wang2003} Y. Wang, S. Ono, Y. Onose, G. Gu, Y. Ando, Y. Tokura, S. Uchida, and N. P. Ong,
Science {\bf 299}, 86 (2003).
\bibitem{tokura} Y. Tokura, H. Takagi, and S. Uchida, Nature (London) {\bf 337}, 345 (1989).
\bibitem{takagi} H. Takagi, S. Uchida, and Y. Tokura, Phys. Rev. Lett. {\bf 62}, 1197 (1989).
\bibitem{jskim} J. S. Kim and D. R. Gaskell, Physica C {\bf 209}, 381 (1993).
\bibitem{kurahashi} K. Kurahashi, H. Matsushita, M. Fujita, and K. Yamada, J. Phys. Soc. Jpn. {\bf 71}, 910 (2002).
\bibitem{fujita1} M. Fujita, T. Kubo, S. Kuroshima, T. Uefuji, K. Kawashima, K. Yamada, I. Watanable, and K. Nagamine,
Phys. Rev. B {\bf 67}, 014514 (2003).
\bibitem{notesdw} Here we call the observed quasi-2D AF scattering spin density wave (SDW) because of its similarity to the quasi-2D AF scattering seen in hole-doped materials \cite{Kimura,Lee}.
\bibitem{daiprl} Pengcheng Dai, H. J. Kang, H. A. Mook, M. Matsuura, J. W. Lynn, Y. Kurita, S. Komiya, and Y. Ando, cond-mat/0501120 (2005).
\bibitem{yamada} K. Yamada, K. Kurahashi, Y. Endoh, R. J. Birgeneau, and G. Shirane,
J. Phys. and Chem. Solids {\bf 60}, 1025 (1999).
\bibitem{yamadaprl} K. Yamada, K. Kurahashi, T. Uefuji, M. Fujita, S. Park,
S.-H. Lee, and Y. Endoh, Phys. Rev. Lett. {\bf 90}, 137004 (2003).
\bibitem{matsuda} M. Matsuda, S. Katano, T. Uefuji, M. Fujita, and K. Yamada, Phys. Rev. B {\bf 66}, 172509 (2002).
\bibitem{Kang} H. J. Kang, Pengcheng Dai, J. W. Lynn, M. Matsuura, J. R.
Thompson, Shou-Cheng Zhang, D. N. Argyriou, Y. Onose, and Y.
Tokura, Nature {\bf 423}, 522 (2003).
\bibitem{Matsuura} M. Matsuura, Pengcheng Dai, H. J. Kang, J. W. Lynn, D. N.
Argyriou, K. Prokes, Y. Onose, and Y. Tokura, Phys. Rev. B {\bf
68}, 144503 (2003).
\bibitem{Chen} Han-Dong Chen, Congjun Wu, and Shou-Cheng Zhang, Phys. Rev.
Lett. {\bf 92}, 107002 (2004).
\bibitem{Mang} P. K. Mang, S. Larochelle, and M. Greven, Nature {\bf 426},
139 (2003).
\bibitem{Kang1} H. J. Kang, Pengcheng Dai, J. W. Lynn, M. Matsuura, J. R.
Thompson, Shou-Cheng Zhang, D. N. Argyriou, Y. Onose, and Y.
Tokura, Nature {\bf 426}, 140 (2003).
\bibitem{Matsuura1} M. Matsuura, Pengcheng Dai, H. J. Kang, J. W. Lynn, D. N. Argyriou,
Y. Onose, and Y. Tokura, Phys. Rev. B {\bf 69}, 104510 (2004).
\bibitem{Mang1}P. K. Mang, S. Larochelle, A. Mehta, O. P. Vajk, A. S. Erickson, L. Lu, W. J. L. Buyers,
A. F. Marshall, K. Prokes, and M. Greven, Phys. Rev. B {\bf 70},
094507 (2004).
\bibitem{boothroyd} A. T. Boothroyd, S. M. Doyle, D. Mck. Paul, and R. Osborn, Phys. Rev. B {\bf 45}, 10075 (1992).
\bibitem{fujitaprl} M. Fujita, M. Matsuda, S. Katano, and K. Yamada, Phys. Rev. Lett. {\bf 93}, 147003(2004).
\bibitem{Lavrov} A. N. Lavrov, H. J. Kang, Y. Kurita, T. Suzuki, Seiki Komiya,
J. W. Lynn, S.-H. Lee, Pengcheng Dai, and Yoichi Ando, Phys. Rev.
Lett. {\bf 92}, 227003 (2004).
\bibitem{Sun} X. F. Sun, Y. Kurita, T. Suzuki, Seiki Komiya, and Yoichi
Ando, Phys. Rev. Lett. {\bf 92}, 047001 (2004).
\bibitem{Jiang} Wu Jiang, J. L. Peng, Z. Y. Li, and R. L. Greene, Phys. Rev. B
{\bf 47}, 8151 (1993).
\bibitem{Radaelli} P. G. Radaelli, J. D. Jorgensen, A. J. Schultz, J. L. Peng, and R. L. Greene,
Phys. Rev. B {\bf 49}, 15322 (1994).
\bibitem{Nath} A. Nath, N. S. Kopelev, V. Chechersky, J. L. Peng, R. L. Greene, O. Boem-hoan, M. I. Larkin, and J. T.
Markert, Science {\bf 265}, 73 (1994).
\bibitem{Riou} G. Riou, P. Richard, S. Jandl, M. Poirier, P. Fournier, V. Nekvasil, S. N. Barilo,
and L. A. Kurnevich, Phys. Rev. B {\bf 69}, 024511 (2004).
\bibitem{Richard} P. Richard, G. Riou, I. Hetel, S. Jandl, M. Poirier, and P. Fournier, Phys. Rev. B
{\bf 70}, 064513 (2004).
\bibitem{sumarlin} I. W. Sumarlin, J. W. Lynn, T. Chattopadhyay, S. N. Sarilo, D. I. Zhigunov,
and J. L. Peng, Phys. Rev. B {\bf 51}, 5824 (1995).
\bibitem{note1} We note that the calibration of the
Cu$^{2+}$ moment is based on normalizing magnetic scattering to
the weak nuclear $(1,1,0)$ Bragg peak without considering
absorption and extinction effects. In addition, since neutrons are
a bulk probe and we can only estimate the moment by assuming an
average ordered moment on all Cu sites. The errors only represent
statistical uncertainties in the calculation.
\bibitem{Skanthakumar} S. Skanthakumar, J. W. Lynn, J. L, Peng,
and Z. Y. Li, J. Appl. Phys. {\bf 73}, 6326 (1993).
\bibitem{Plakhty} V. P. Plakhty, S. V. Maleyev, S. V. Gavrilov, F. Bourdarot, S.
Pouget, and S. N. Barilo, Europhys. Lett. {\bf 61}, 534 (2003).
\bibitem{li} S. L. Li, S. D. Wilson, D. Mandrus, B. R. Zhao, Y. Onose, Y. Tokura, and Pengcheng Dai,
Phys. Rev. B {\bf 71}, 054505 (2005).
\bibitem{note2} Whether the 3D AF phase coexists with superconductivity macroscopically or
microscopically in NCCO is still an open question. In addition, it
is not clear why optimally doped NCCO has 3D AF order in
coexistence with superconductivity while similar PLCCO does not.
\bibitem{khaykovich3} B. Khaykovich, S. Wakimoto, R. J. Birgeneau, M. A. Kastner, Y. S. Lee, P. Smeibidl, P. Vorderwisch, and K. Yamada, cond-mat/0411355 (unpublished).
\end{references}
\end{document}